\documentclass[11pt,a4paper]{article}
\pdfoutput=1
\usepackage{jheppub}
\usepackage{graphicx}
\usepackage{lipsum}
\usepackage{youngtab}
\usepackage{tikz}
\usepackage{tikz-cd,quiver}
\usepackage{braids}
\usepackage{amsmath,amsfonts}
\usepackage{slashed}
\usepackage{enumitem}
\usepackage{chngpage}
\usepackage[all]{hypcap}
\allowdisplaybreaks

\usepackage{amsthm}

\theoremstyle{definition}

\def\Z{\mathbb{Z}}
\def\mod{~\text{mod~}}
\graphicspath{ {./Figs/} }
\usepackage{xspace}

\usepackage{amscd,amsxtra,mathrsfs,graphics,graphicx,amsthm,epsfig,bm,longtable,float,color,tikz,mathtools,xfrac,footnote,rotating,lscape,wrapfig}
\usepackage{simpler-wick}
\usepackage[makeroom]{cancel}
\usepackage{bm}
\usepackage{physics}
\usepackage{caption,subcaption}
\usetikzlibrary{arrows}
\usepackage{empheq}
\usepackage{array}   
\newcolumntype{C}{>{$}c<{$}} 

\def\nn{\nonumber}

\def\SU{\mathrm{SU}}
\def\su{\mathfrak{su}}

\def\U{\mathrm{U}(1)}

\def\SL{\mathrm{SL}}
\def\Spin{\mathrm{Spin}}

\def\UU{\text{U}}
\def\u{\mathfrak{u}}
\def\g{\mathfrak{g}}

\def\diag{\text{~diag}}
\def\ii{\text{i}}

\def\ee{\text{e}}

\def\F{\mathbb{F}}

\def\rbord{\tilde{\Omega}^{\Spin}}

\def\hom{\text{~Hom}}
\def\ext{\text{Ext}}

\def\sq{\text{~Sq}}

\title{Anomalies of non-Abelian finite groups via
    cobordism}
\date{}
\author[a]{Joe Davighi,}
\author[b]{Ben Gripaios,}
\author[c]{and Nakarin Lohitsiri}

\affiliation[a]{Physics Institute, University of Zurich}
\emailAdd{joedavighi@gmail.com}
\affiliation[b]{Cavendish Laboratory, University of Cambridge}
\emailAdd{gripaios@hep.phy.cam.ac.uk}
\affiliation[c]{Department of Mathematical Sciences, Durham University}
\emailAdd{nakarin.lohitsiri@durham.ac.uk}

\abstract{
We use cobordism theory to analyse anomalies of finite
non-abelian symmetries in 4 spacetime dimensions. 
By applying the method of `anomaly interplay', which uses
  functoriality of cobordism and naturality of the $\eta$-invariant to relate
anomalies in a group of interest to anomalies in other (finite or
compact Lie) groups,
we derive the anomaly for every  representation
in many examples motivated by flavour physics, including
$S_3$, $A_4$, $Q_8$, and $\mathrm{SL}(2,\F_3)$. 

In the case of finite abelian groups, it is well known that
anomalies can be `truncated' in a
way that has no effect on low-energy physics, by means of a group extension.
We extend this idea to non-abelian symmetries. We show, for example,
that a system with $A_4$ symmetry can be rendered anomaly-free, with only one-third as many fermions as na\"ively required, by passing to a larger symmetry. As another example, we find that a well-known model of quark and lepton masses utilising the $\SL(2,\F_3)$ symmetry is anomalous, but that the anomaly can be cancelled by enlarging the symmetry to a $\Z/3$ extension of $\SL(2,\F_3)$.
}

\begin{document}
\maketitle

\section{Introduction}
\label{sec:intro}

Chiral symmetries of massless fermions typically exhibit anomalies,
which play an important role in quantum field theory (QFT). If the
chiral symmetry is gauged, then anomalies must cancel, modulo a coupling to a topological
quantum field theory (TQFT). If the chiral symmetry is global, then
there can be mixed `ABJ anomalies'~\cite{Bell:1969ts,Adler:1969gk}
that render the classical symmetry broken, and there can also be 't Hooft
anomalies~\cite{tHooft:1979rat} that prevents us from coupling background gauge fields to global symmetries.
Due to their topological nature, 't
Hooft anomalies are not renormalised and so provide powerful tools for analysing QFTs, especially in r\'egimes of strong coupling.

It is now known that all such chiral fermion anomalies are captured by
cobordism
groups~\cite{Witten:1985xe,Dai:1994kq,Kapustin:2014dxa,Witten:2015aba,Freed:2016rqq,Witten:2019bou},
where `cobordism' is a particular cohomology theory.\footnote{For us, the word `bordism' will always refer to a covariant functor (that was once upon a time introduced as `cobordism'), while we reserve the word `cobordism' to denote its contravariant dual. Roughly speaking, the elements in a bordism group are equivalence classes of manifolds, equipped with certain structures, where two manifolds are considered equivalent if there exists a manifold in one dimension higher joining them. By cobordism groups, we mean the (shifted) Anderson dual to bordism, or in a special case the Pontryagin dual.} Given a symmetry
type,\footnote{The `symmetry type', defined in the technical sense of
  Definition 2.4 in Freed and Hopkins' Ref.~\cite{Freed:2016rqq},
  includes not only the internal symmetry but also the spacetime
  symmetry, allowing for possible quotients between the two. As is
  usual in the study of anomalies, we assume Euclidean signature
  throughout.  } an appropriate cobordism group classifies both
perturbative (`local') anomalies, {\em i.e.} those computed via
one-loop Feynman diagrams, and also non-perturbative
(`global') anomalies~\cite{Witten:1982fp} in that symmetry. While
local anomalies are easily analysed by more pedestrian means, in the
case of global anomalies the rigorous cobordism classification
supersedes previous arguments based on
homotopy~\cite{Davighi:2020kok}, and allows anomalies to be analysed
systematically using the tools of algebraic topology. 

This cobordism
classification of anomalies has been used in many contexts in recent
years, with applications in both high energy physics (including string
theory) and condensed matter physics. On the high energy side, applications include the elucidation of an order 16 anomaly in $\frac{\mathrm{Spin} \times \Z/4}{\Z/2}$
symmetry~\cite{Tachikawa:2018njr} and related
anomalies~\cite{Wan:2018bns}, a new $SU(2)$ anomaly for non-spin spacetimes~\cite{Wang:2018qoy}, anomalies (and their absence) in the
Standard Model (SM) and Beyond the
SM~\cite{Freed:2006mx,Garcia-Etxebarria:2018ajm,Davighi:2019rcd,Wan:2020ynf,Davighi:2022fer,Wang:2022eag},
examples in 6d~\cite{Lee:2020ewl,Davighi:2020kok} and
8d~\cite{Garcia-Etxebarria:2017crf,Lee:2022spd}, anomalies in duality groups
\cite{Seiberg:2018ntt,Hsieh:2019iba}, anomaly cancellation in heterotic string theory~\cite{Tachikawa:2021mvw,Tachikawa:2021mby}, and a newly discovered anomaly
in Type IIB string theory~\cite{Debray:2021vob}. 

In this paper, we apply the rigorous cobordism classification of anomalies to the case of finite non-abelian symmetries in 4d. Through anomaly inflow, it has been shown that fermion anomalies are given by the exponentiated $\eta$-invariant evaluated on a manifold in one dimension higher~\cite{Witten:2019bou}. For a   finite group $G$, in 4d, there are no perturbative anomalies and so
  the exponentiated $\eta$-invariant becomes a 5d bordism invariant. Thus, if fermions are defined using an ordinary  spin structure, then the possible anomalies are elements of the (abelian)
    group of homomorphisms from the bordism group
    $\Omega_5^\mathrm{Spin}(BG)$ to $\U$, given by the exponentiated $\eta$-invariant. The first step in characterising the anomalies is thus to find that bordism group.
In the abelian case, the bordism groups were explored by Hsieh in
Ref.~\cite{Hsieh:2018ifc} (see also~\cite{Garcia-Etxebarria:2018ajm}), and a smattering of non-abelian finite
group anomalies have been studied in the literature.\footnote{Examples
  include quaternionic groups in 6d~\cite{Garcia-Etxebarria:2018ajm} and dihedral groups that
  mix with spacetime symmetry \cite{Debray:2021rik}.}  

We here compute
$\Omega_d^\mathrm{Spin}(BG)$, in degrees $d=1, \dots, 6$, for a variety of
non-abelian finite groups (see Table~\ref{tab:bord-results}). The examples that we choose to showcase our methods are not random, but chosen
because of
  their applications to flavour physics,
where the groups $S_3$, $A_4$, $Q_8$, and $\SL(2,\F_3)$ have been
widely used in explanations of the quark flavour hierarchies and/or
neutrino mass and mixing
data~\cite{Frampton:1995fta,Aranda:1999kc,Ma:2001dn,Altarelli:2005yp,Carr:2007qw,Feruglio:2007uu}. 

We then proceed to compute the anomalies, by determining the exponentiated $\eta$-invariant for each irreducible representation (henceforth {\em irrep}) of $G$, for the chosen examples listed in Table~\ref{tab:bord-results} in 4d.
We use a variety of tools to perform this
task completely.
We find that $S_3$
and $Q_8$ are completely anomaly free, while $A_4$ and $\SL(2,\F_3)$
suffer from anomalies of order 9.\footnote{
By the {\em order} of an anomaly, we mean the order of the image of the exponentiated $\eta$-invariant in the appropriate cobordism group (which, for us, is always a finite group). This notion will be made precise in \S \ref{sec:cobord-class}.}
However, the moral of the story is that the anomalies should be computable for any group of interest, if one
  knows enough tricks of algebraic topology.

The main tool we use for deriving these anomalies is the fact that the exponentiated $\eta$-invariant is a natural transformation between two appropriate functors (see~\S \ref{sec:interplay}), 
through which the (unknown) anomalies in symmetry $G$ are related to
(known) anomalies in a different group $H$, where $G$ and $H$ are
related by a homomorphism. This mapping between anomalies in different symmetry types, dubbed `anomaly
interplay'~\cite{Davighi:2020bvi,Davighi:2020uab,Grigoletto:2021zyv,Grigoletto:2021oho}, is particularly useful when there are
non-trivial maps going in both directions between $G$ and $H$; for
example, this occurs when $G$ is a semi-direct
product $K\rtimes H$, as is the case in several of our examples {\em
  e.g.} $\SL(2,\F_3) \cong Q_8 \rtimes \Z/3$.

In the context of particle physics, there is a subtlety pertaining to
finite group symmetries that is not seen for
connected Lie groups. The subtlety, which was first appreciated in the
context of $\Z/k$ symmetries by Banks and Dine~\cite{Banks:1991xj}
(see~\cite{Hsieh:2018ifc} for a discussion from the bordism
perspective), is that anomalies in a finite symmetry $G$ are not
`invariant'
under extensions of $G$, where such extensions cannot necessarily be
distinguished from $G$ by the low-energy physicist. In particular, a representation $\textbf{r}$ of $G$ whose exponentiated $\eta$-invariant is non-trivial can pull back to a representation of a group extension $H \twoheadrightarrow G$ with a trivial exponentiated $\eta$-invariant. For
example, take $G=\Z/3$, and let the (reducible) representation $\bf{r}$ denote three copies of the fundamental representation of $\Z/3$. While the representation $\bf{r}$ has an anomaly $\exp (2\pi \ii/3)$, meaning that na\"ively three such fermions would be required to be anomaly-free, 
the representation $\bf{r}$ nevertheless pulls back to an anomaly-free representation of the extension $\Z/9 \twoheadrightarrow \Z/3$. 

The final goal of this paper is to extend this idea of `truncating'
 finite anomalies by a group extension to the case of finite non-abelian groups. In both cases where we find non-trivial anomalies of order 9, namely for $G=A_4$ and $G=\SL(2,\F_3)$, we show that the anomaly is truncated to an order 3 anomaly by a non-trivial extension of $G$ by $\Z/3$. The anomalies that are left are consistent with the recent Ref.~\cite{Gripaios:2022vvc} (see also \cite{Talbert:2018nkq, Kobayashi:2021xfs}), which analysed the `linear part' of finite gauge anomalies using character theory.

There are some new mathematical results in this paper, including: 
\begin{itemize}
\item Calculation of the odd torsion for $\Omega^{\Spin}_d(B\SL(2,\mathbb{F}_3))$ for $d\le 6$ (Appendix \ref{app:SL2F3});
\item Proof that $\Omega^{\Spin}_5(BD_{2n})=0$ for any $n\in \Z$ (Appendix \ref{app:D2n});
\item Calculation of $\Omega^{\Spin}_d(BD_{2n})$ for $d\leq 5$ when $n$ is a power of $2$ (Appendix \ref{app:D2n});
\item Computations of the exponentiated $\eta$-invariant for all complex (and pseudo-real) irreps of $A_4$ and $\SL(2,\mathbb{F}_3)$ in \S\S~\ref{sec:A4-4d} and \ref{sec:SL23}.
\end{itemize}

The structure of the paper is as follows.
In \S\ref{sec:cobord-class} we review the precise connection between chiral fermion anomalies and $\eta$-invariants and the consequent classification of anomalies using cobordism, specialising to the case of finite groups. We review the tools of anomaly interplay on which we rely. In \S\ref{sec:bord-table} we record all our spin-bordism computations, for a selection of finite non-abelian groups in degrees 1 through 6 -- the calculations themselves are collected in Appendix~\ref{app:bordism-results}.
In \S\ref{sec:main} we use anomaly interplay to compute $\eta$-invariants for these groups, and thus derive complete information about anomalies. In \S\ref{sec:truncation} we formally introduce the notion of `anomaly truncation' by finite group extensions, and show how this works out in our two non-trivial examples, before summarising in \S \ref{sec:summary}.

\section{The cobordism classification of anomalies for finite groups}
\label{sec:cobord-class}

By now it is well-established that fermionic anomalies are captured by (co)bordism groups. We begin by briefly reviewing why this is the case.
Consider a system of massless chiral fermions $\{\psi\}$ on a
$d$-dimensional closed manifold $\Sigma_d$, with a metric $g$, spin
structure $s$, and principal $G$-bundle with connection $A$. For now, $G$ may
be either a compact Lie group or a finite group.

Recently, Witten and Yonekura proved~\cite{Witten:2019bou} a
long-suspected formula~\cite{Witten:1985xe,Dai:1994kq,Witten:2015aba},
which is that the phase of the fermionic partition function obtained
by integrating over $\{\psi\}$ is equal to the exponentiated
$\eta$-invariant of Atiyah, Patodi and
Singer~\cite{Atiyah:1975jf,Atiyah:1976jg,Atiyah:1976qjr}, evaluated on
a manifold $X_{d+1}$ whose boundary is $\Sigma_d$ and to which
all the structure extends.\footnote{If the relevant
  bordism group in degree $d$ is non-vanishing, then there are manifolds for which no extension exists. A 4d example is the K3 manifold, for which there is no extension of the spin structure to a 5-manifold bounded by K3. For such non-nullbordant spacetimes, the partition function suffers from an ambiguity that corresponds to assigning a generalised theta angle to each generator of the bordism group $\Omega_d$~\cite{Freed:2004yc}. } 
We write $Z_\psi[\Sigma, (A, g, s)] = |Z_\psi| \exp \left( 2\pi \ii \eta(X_{d+1})\right)$.

For this partition function to make sense as a description of
local physics in $d$-dimensions, it should not depend on the
choice of auxiliary manifold $X_{d+1}$, nor on the particular choice
of extension of structure thereto. Thus, if we had chosen another
such extension $X^\prime_{d+1}$ with structure, then we must have
$\exp \left( 2\pi \ii \eta(X^\prime_{d+1})\right)= \exp \left( 2\pi \ii
  \eta(X_{d+1})\right)$. Now, since $X$ and $X^\prime$ share a
boundary on which the structures agree, one can glue $X_{d+1}$ and
$-X^\prime_{d+1}$, where the minus sign denotes orientation reversal,
to form a closed $(d+1)$-manifold $\overline{X}_{d+1}$. The
exponentiated $\eta$-invariant obeys a gluing law~\cite{Dai:1994kq},
which implies our previous condition is equivalent to requiring $\exp
\left( 2\pi \ii \eta(\overline{X}_{d+1})\right) = 1$. To be fully
consistent with locality, one thus requires that 
\begin{equation} \label{eq:locality}
\exp \left( 2\pi \ii \eta(\overline{X}_{d+1})\right) = 1
\end{equation}
for {\em all} closed $(d+1)$-manifolds $\overline{X}$ with structure. 

The condition (\ref{eq:locality}) implies that the theory is
anomaly-free in the more traditional sense. This is because, given any
gauge transformation $A \to A^g$ of the background connection (possibly
combined with a diffeomorphism $\varphi$ of $\Sigma_d$), one can construct a
mapping torus $T^g_{d+1} \cong \Sigma_d \times_g S_1$ by taking a cylinder and
gluing the ends using the gauge transformation (plus
diffeomorphism). By cutting and gluing, one sees that the variation of
the partition function is given by
$Z[\Sigma, (A, g, s)]/Z[\varphi(\Sigma), \varphi(A^g, g, s)] = \exp \left( 2\pi \ii \eta(T^g_{d+1})
\right)$, where we have used the fact that the modulus of the
partition is necessarily anomaly-free. The mapping torus $T^g_{d+1}$
is a closed $(d+1)$-manifold with structure $(A, g, s)$, on which the
exponentiated $\eta$-invariant is trivial by (\ref{eq:locality}). Thus,
the locality condition implies complete anomaly-freedom, from both
perturbative and non-perturbative anomalies in the traditional sense,
but is much stronger.\footnote{Such a condition is also thought to be
  needed ultimately for consistently quantizing gravity; see {\em
    e.g.}~\cite{McNamara:2019rup,Ooguri:2020sua} for related
  arguments.}

In this paper we are interested in the special case of theories with
symmetry group $G$ being a {\em finite group}, for which the analysis
of anomalies simplifies. Let us further assume for now that
$d=4$.\footnote{ In other dimensions, {\em e.g.} in 2d or 6d, the
  situation is complicated somewhat due to the presence of pure
  gravitational anomalies. In practice, however, these pose no
  significant issue for analysing anomalies, because one can trivially
  cancel such pure gravity anomalies by adding enough neutral fermions
  in the opposite
  chirality. } In this scenario, the fermion anomaly given by the exponentiated
  $\eta$-invariant\footnote{Like any anomaly theory, the exponentiated
    $\eta$-invariant for a more general symmetry structure is an
    example of a reflection positive invertible quantum field
    theory. Much is known mathematically about the space of such
    theories -- for example, in the case of anomaly theories that are
    moreover topological, they are rigorously classified in terms of
    stable homotopy~\cite{Freed:2016rqq}.} becomes a bordism
  invariant. This can be easily seen by evaluating the exponentiated
  $\eta$-invariant on a closed manifold $\overline{X_5}$ that bounds
  $Y_6$. The Atiyah--Patodi--Singer (APS) index theorem tells us that
\begin{equation}
\label{eq:APS-index-thm}
\exp \left( 2\pi\ii \eta(X_5) \right) = \exp \left( 2\pi\ii \int_{Y_6}\Phi_6 \right),
\end{equation}
where $\Phi_6$, the anomaly polynomial, is a closed $6$-form with integer
periods built from the curvature of $Y_6$ and the field strength of
the connection $A$. For finite $G$, $A$ is flat and consequently
$\Phi_6$ vanishes.\footnote{Because we are in $d=4$, there is no `pure
  gravitational anomaly' terms in the anomaly polynomial that do not
  depend on the field strength of $A$.}  Thus, the exponentiated
$\eta$-invariant evaluated on a pair of bordant manifolds $X_5$ and
$X_5^{\prime}$ must be equal, since
\begin{equation}
\frac{\exp \left( 2\pi\ii \eta(X_5) \right)}{\exp \left( 2\pi\ii \eta(X_5^{\prime}) \right)} = \exp \left( 2\pi\ii \eta(X_5\sqcup (-X_5^{\prime})) \right) = 1,
\end{equation}
where we used the fact that $X_5\sqcup (-X_5^{\prime})$ is a boundary to get the last equality. Therefore, the exponentiated $\eta$-invariant defines an element of the group of homomorphisms from the bordism group $\Omega^{\Spin}_5(BG)$ to $\U$, which we denote
\begin{equation}
  \label{eq:Pontryagin-dual}
  \mho^6_{\Spin}(BG) := \hom \left( \Omega^{\Spin}_5(BG), \U \right)\, ,
\end{equation}
adopting the notation of {\em e.g.}~\cite{Kaidi:2019tyf}.
This abelian group is pure torsion,\footnote{This can be seen from looking at the $E^2$ page of the
    Atiyah--Hirzebruch spectral sequence for the spin bordism group of
    $BG$ associated to the fibration $\text{pt} \to BG \to BG$:
\begin{equation}
E^2_{p,q} = H_p(BG;\Omega^{\Spin}_q(\text{pt})) \Rightarrow \Omega^{\Spin}_{p+q}(BG).
\end{equation}
On the diagonal $p+q=5$, the only possibly non-trivial entries are of
the form $H_i(BG;A)$ where $i>0$ and $A=\Z$ or $\Z/2$. These homology
groups are finite when $G$ is finite. Since
$\abs{\Omega^{\Spin}_5(BG)} \leq \abs{\bigoplus_{p+q=5}E^2_{p,q}}$, we obtain
the desired result.} implying that the anomaly is of finite order, as is always the case for a global anomaly. In
the rest of the paper, we call $\mho_\Spin^6(BG)$ a {\em
  cobordism group}.\footnote{This agrees with a more general (and more widely used) usage
  of the term cobordism  to mean the (shifted) Anderson dual
  of the bordism group \cite{Freed:2014eja,Freed:2016rqq,Davighi:2020uab}. When $G$ is finite and $d=4$, as we consider
  here, the two coincide. }

Clearly, we can specify
the exponentiated $\eta$-invariant  by giving its values of a set of generators of
$\Omega_{5}^\Spin(BG)$ for a single
representation from each equivalence class of irreps of $G$.
From there, the complete set of conditions  for anomaly cancellation in a 4d finite $G$ gauge theory can be obtained. While this is in general a daunting task, for the abelian case, in which $G = \prod_{i} \Z/k$ is a product of cyclic groups, this task has been completed in Ref.~\cite{Hsieh:2018ifc}.  One goal of this paper is to extend these computations to some non-abelian finite groups, such as $S_3$, $A_4$, $Q_8$, and $\SL(2,\mathbb{F}_3)$.

\subsection{A tool: the Frobenius--Schur indicator}

One elementary fact that we can easily leverage is that, in 4d, chiral fermions transforming in real representations are anomaly-free.
For finite groups, it is especially straightforward to check whether a representation $\bf{r}$ is real simply from its character $\chi(g)$.\footnote{For compact Lie groups things are only slightly more involved, with the Frobenius--Schur indicator then being given by the integral $\iota_{\bf r} = \int_{g\in G} \chi(g^2) d\mu$, where $d\mu$ is the Haar measure normalised such that $\int d\mu = 1$. 
} To wit, one computes the Frobenius--Schur indicator, 
\begin{equation} \label{eq:FS}
\iota_{\bf r} := \frac{1}{|G|}\sum_{g\in G} \chi(g^2) \, ,
\end{equation}
which equals $1$, $0$, or $-1$ when ${\bf r}$ is a real, complex, or pseudo-real irrep, respectively.
To compute the anomaly for a complex or pseudo-real representation,
in 4d, requires less trivial methods.

\subsection{A strategy: anomaly interplay} \label{sec:interplay}

The main strategy that we use to calculate anomalies is to exploit `anomaly interplay'. The idea is to pullback anomalies from one symmetry type to another, by viewing the exponentiated $\eta$-invariant as a natural transformation between contravariant functors. (Here we use the term `anomaly' and the exponentiated $\eta$-invariant interchangeably.) Physics-wise, this harks back to work of Elitzur and Nair~\cite{ELITZUR1984205} who, following Witten~\cite{Witten:1983tw}, showed that the 4d $SU(2)$ anomaly can be derived by `pulling back' a local anomaly from {\em e.g.} $SU(3)$. Anomaly interplay is rather trivially defined with a little category theory, which we now introduce before continuing.

  Let $\mathrm{FinGp}$ be the usual category of finite groups and let
  $\mathrm{Ab}$ be the usual category of abelian groups. Then
  $\mho^6_{\Spin}(B\cdot)$ defines a contravariant functor
  $\mathrm{FinGp}\to \mathrm{Ab}$. A second contravariant functor
  $\mathrm{FinGp}\to \mathrm{Ab}$ is given by the representation ring
  $RU(\cdot)$, where we forget the ring multiplication, regarding
  $RU(G)$ just as an abelian group. Concretely, $RU(G)$ is a free
  abelian group with basis given by (isomorphism classes of)
  irreps of $G$ and sum given by direct sum of
  representations. Writing $\Z \left\langle \textbf{r}_i \right\rangle$ for the free abelian group generated by an irrep $\textbf{r}_i$, we can write $RU(G)$ explicitly as
\begin{equation} \label{eq:RUG-def}
RU(G) = \bigoplus_i \Z \left\langle \textbf{r}_i \right\rangle,
\end{equation}
where $i$ runs through all irreps of $G$. We emphasise that, for our purposes, (\ref{eq:RUG-def}) should be thought of as a free abelian group, and we forget its ring structure. This permits us to consider subgroups thereof.

The exponentiated $\eta$-invariant 
is a natural transformation from the functor $RU(\cdot)$ to the functor $\mho^6_{\Spin}(B\cdot)$ \cite{Grigoletto:2021zyv,Grigoletto:2021oho}.
This means that, given a homomorphism $f:G\to K$ between finite groups, there are pullbacks
  induced on both these functors,
\begin{align}
    RU(f):\, & RU(K) \to RU(G)\,, \\
    \mho(f):\, & \mho^6_{\Spin}(BK) \to \mho^6_{\Spin}(BG)\,,
\end{align}
such that the diagram
\begin{equation} \label{eq:ROm_square}
\begin{tikzcd}
{RU(K)} & {RU(G)} \\
{\mho^6_{\text{Spin}}(BK)} & {\mho^6_{\text{Spin}}(BG)}
\arrow["{RU(f)}"', from=1-1, to=1-2]
\arrow["{\mho(f)}"', from=2-1, to=2-2]
\arrow["{\exp (2\pi \text{i}\eta)_K}"', from=1-1, to=2-1]
\arrow["{\exp (2\pi\text{i}\eta)_G}", from=1-2, to=2-2]
\end{tikzcd}\, 
\end{equation}
commutes (where we use the usual subscript notation to denote a component of the natural transformation $\exp (2\pi \text{i}\eta)$). 
By `anomaly interplay' we simply refer to the constraints placed on anomalies of $G$ that follow from this commutative diagram.
This version of anomaly interplay, in which both groups involved are finite, has been applied to the study of
anomalies in two-dimensions in Ref. \cite{Grigoletto:2021zyv,Grigoletto:2021oho}.\footnote{Another version of  anomaly interplay between anomalies of finite groups and anomalies of compact Lie groups has been used in Refs. \cite{Davighi:2020bvi,Davighi:2020uab}. In fact, at one point in this paper (\S \ref{sec:interplay-with-SM}) we will need to go outside the framework we have set up, and consider a compact Lie group, namely $\UU(2)$, instead of a finite group. But since we do so for a representation of $\UU(2)$ which has no local anomalies, very similar arguments apply.}

One particular trick we will frequently employ is to pullback whole short exact sequences that relate a non-abelian group $G$ to {\em abelian} finite groups. For example, we are guaranteed a short exact sequence defining the `abelianisation' of $G$:
\begin{equation}
0 \to G^\prime \to G \to G/G^\prime \to 0\, .
\end{equation}
Here, $G^\prime$ is the derived subgroup of $G$, {\em i.e.} the normal subgroup of $G$ generated by commutators $[g,h]:=ghg^{-1}h^{-1}$. The quotient group $G/G^\prime$ is necessarily abelian, and is called the abelianisation of $G$.
By pulling back such sequences, particularly in cases where they split, often allows us to completely derive the anomalies in $G$ from known building blocks, {\em i.e.} from anomalies in the abelian case.

Before we begin in earnest, we introduce a few more notions that will be useful in what follows.
Firstly, the {\em order of the anomaly of a representation $\textbf{r}$} is the order of the image of $\textbf{r}$ under the exponentiated $\eta$-invariant. 
We can also define a slightly more general notion, which is the {\em order of the anomaly of a finite group $G$}, to be simply the order of the image of the exponentiated $\eta$-invariant in $\mho^6_\Spin(BG)$.
Lastly, we define the {\em subgroup of anomaly-free representations} $RU_0(G)$ to be
\begin{equation}
RU_0(G) := \ker \left( \exp(2\pi\ii \eta) \right) \subset RU(G)
\end{equation}
for any finite group $G$. This notion will be especially useful when we come to discuss the `truncation' of finite group anomalies in \S \ref{sec:truncation}.

\section{Bordism group results} \label{sec:bord-table}

To compute the relevant bordism groups, we rely on standard spectral sequence techniques, using a combination of the Atiyah--Hirzebruch spectral sequence (AHSS) and the Adams spectral sequence (ASS).
We tabulate our results for the (reduced) spin-bordism groups of $BG$ for various non-abelian finite groups $G$ in Table~\ref{tab:bord-results}. 

The computations themselves, some of which can be found in existing literature, are recorded in Appendix \ref{app:bordism-results}. These bordism computations provide the platform from which we systematically, and comprehensively, study anomalies in 4d for various finite $G$ symmetries.

\begin{table}[h]
\begin{adjustwidth}{-.5in}{-.5in}  
  \centering
  \begin{tabular}{|c|cccccc|}
    \hline
    $p$ & $1$ & $2$ & $3$ & $4$ & $5$ & $6$\\
    \hline
    $\rbord_p(BS_3)$ & $\Z/2$ & $\Z/2$ & $\Z/8\times \Z/3$ & $0$ & $0$ & $0$\\
    $\rbord_p(BA_4)$ & $\Z/3$ & $\Z/2$ & $\Z/12$ & $0$ & $\Z/9$ & $\Z/2$\\
    $\rbord_p(BQ_8)$ & $(\Z/2)^2$ & $(\Z/2)^2$ & $\Z/8 \times (\Z/4)^2$ & $\Z/2$ & $(\Z/2)^2$ & $\Z/2$ \\
    $\rbord_p(B\text{SL}(2, \F_3))$ & $\Z/3$ & $0$ & $\Z/8 \times \Z/3$ & $\Z/2$ & $(\Z/2)^2\times \Z/9$ & $\Z/2$\\
    $\rbord_p(BD_{2n})$, \text{~$n$ odd} & $\Z/2$ & $\Z/2$ & $\Z/8\times \Z/n$ & $0$ & $0$ & $0$\\
    $\rbord_p(BD_{2n})$, \text{~$n=2^{k+1}$ } & $(\Z/2)^2$ & $(\Z/2)^3$ & $(\Z/8)^2\times \Z/(2n)$ & $(\Z/2)^2$ & $0$ & $\Z/2$\\
    \hline
  \end{tabular}
\end{adjustwidth}
  \caption{Reduced spin-bordism groups in degrees 1 through 6 of $BG$, for various finite non-abelian groups $G$. The full ({\em i.e.} unreduced) spin bordism groups are obtained as $\Omega_p^{\mathrm{Spin}}(BG) \cong \rbord_p(BG) \oplus \Omega_p^{\mathrm{Spin}}(\mathrm{pt})$.
}
  \label{tab:bord-results}
\end{table}

\section{Computing finite group anomalies by anomaly interplay} \label{sec:main}

\subsection{$G=A_4$} \label{sec:A4-4d}

For our first example we consider $G=A_4$, the group of even permutations of four elements, equivalently the orientation-preserving symmetries of a regular tetrahedron. This  group has often been used in flavour model building, particularly for controlling neutrino masses and mixing angles ({\em e.g.}~\cite{Ma:2001dn,Altarelli:2005yp}).

There are three one-dimensional irreps  and one
three-dimensional irrep of $A_4$. We denote these irreps by their
dimensions, as $\textbf{1}$, $\textbf{1}^{\prime}$,
$\overline{\textbf{1}^{\prime}}$, and $\textbf{3}$, respectively. The
character table (which can be obtained from {\em e.g.} the computer program \texttt{GAP} \cite{GAP4} or from Ref. \cite{Fulton2004})
is given in Table \ref{tab:A4-char}, where we use a
presentation of $A_4$ with two generators $\sigma$ and $\tau$ of order 2 and 3 satisfying $(\tau\sigma)^3=1$. The representation ring $RU(A_4)$ for the group $A_4$ can then be written explicitly as
  \begin{equation}
    \label{eq:RUA4}
RU(A_4) = \Z\, \left<\textbf{1}\right> \oplus \Z\, \left<\textbf{1}^{\prime}\right> \oplus \Z\,\left<\overline{\textbf{1}^{\prime}}\right> \oplus \Z\, \left<\textbf{3}\right> .
\end{equation}
\begin{table}[h]
  \centering
  \begin{tabular}{c|cccc}
    & $1$ & $4$ & $4$ & $3$\\
    $A_4$ & $1$ & $\sigma$ & $\sigma^2$ & $\tau$\\
    \hline
    $\textbf{1}$ & $1$ & $1$ &$1$ & $1$\\
    $\textbf{1}^{\prime}$ & $1$ & $\omega$ & $\omega^2$ & $1$\\
    $\overline{\textbf{1}^{\prime}}$ & $1$ & $\omega^2$ & $\omega$ & $1$\\
    $\textbf{3}$ & $3$ & $0$ & $0$ & $-1$ 
  \end{tabular}
  \caption{Character table of $A_4$, the alternating group of order
    $4$. Conjugacy classes are labelled by a representative element with the numbers of elements in each class atop. $\omega$ is a cube root of unity.}
  \label{tab:A4-char}
\end{table}

In Appendix~\ref{app:A4} we record the spin bordism groups of $BA_4$, taken from~\cite{bruner2010connective}. Of relevance to 4d anomalies, we have
$\Omega_5^\Spin(BA_4) \cong \Z/9$.
Since there are no pure gravitational anomalies in 4d, the cobordism group that classifies anomalies is simply
\begin{equation}
\mho^6_\Spin(BA_4) \cong \Omega_5^\Spin(BA_4) \cong \Z/9\, 
\end{equation}
also.
In this Section we calculate the anomalies for $G=A_4$ in 4d, using anomaly interplay. We will compute the anomaly associated with every irrep of $A_4$, and from there write down the general condition for $A_4$ anomaly cancellation.

To begin, recall that the derived subgroup of $A_4$ is the Klein four-group, $A_4^\prime = \Z/2 \times \Z/2$. 
The abelianisation short exact sequence is
\begin{equation}
  \label{eq:A4-SES}
0 \to \Z/2 \times \Z/2 \to A_4 \to \Z/3 \to 0\, .
\end{equation}
This sequence moreover splits, meaning that $A_4$
can be realised as a semi-direct product
\begin{equation}
A_4 = \left( \Z/2\times \Z/2 \right) \rtimes \Z/3\, ,
\end{equation}
so there is a map $j: \Z/3 \to A_4$ that embeds $\Z/3$ as a (non-normal)
subgroup of $A_4$.  
We therefore obtain the following commutative
diagram, where the horizontal row is exact,
\begin{equation}
  \label{eq:A4-commutative-diag}
\begin{tikzcd}
	0 & {\mathbb{Z}/2\times \mathbb{Z}/2} & {A_4} & {\mathbb{Z}/3} & 0 \\
	& {} & {\mathbb{Z}/3}
	\arrow[from=1-1, to=1-2]
	\arrow["i", from=1-2, to=1-3]
	\arrow["\pi", from=1-3, to=1-4]
	\arrow[from=1-4, to=1-5]
	\arrow["j", from=2-3, to=1-3]
	\arrow[Rightarrow, no head, from=2-3, to=1-4]
\end{tikzcd}
\end{equation}    
For the other groups involved in this diagram, the relevant cobordism groups are $\mho_\Spin^6(B(\Z/2\times \Z/2)) \cong \Omega^{\Spin}_5(B(\Z/2\times \Z/2))=0$~\cite{YuPHD:1995} (see Appendix~\ref{app:Z2Z2}), and $\mho_\Spin^6(B\Z/3) \cong \Omega^{\Spin}_5(B\Z/3) \cong \Z/9$.

Applying the representation ring functor and the cobordism functor to the right-hand triangle in (\ref{eq:A4-commutative-diag}), and relating the two via the $\exp(2\pi \ii \eta)$ natural transformation, we get the following commutative diagram:
\begin{equation}\label{eq:A4-4d-interplay}
\begin{tikzcd}
	{} & {RU(A_4)} && {RU(\mathbb{Z}/3)} \\
	&& {RU(\mathbb{Z}/3)} \\
	& {\mho^6_{\mathrm{Spin}}(BA_4)\cong\Z/9} && {\mho^6_{\mathrm{Spin}}(B\mathbb{Z}/3)\cong\Z/9} \\
	&& {\mho^6_{\mathrm{Spin}}(B\mathbb{Z}/3)\cong\Z/9}
	\arrow["{\qquad\pi^\ast}"', from=3-4, to=3-2]
	\arrow["{j^\ast}", from=3-2, to=4-3]
	\arrow[Rightarrow, no head, from=4-3, to=3-4]
	\arrow[Rightarrow, no head, from=2-3, to=1-4]
	\arrow["{j^\ast}", from=1-2, to=2-3]
	\arrow["{\pi^\ast}"', from=1-4, to=1-2]
	\arrow[color={rgb,255:red,153;green,92;blue,214}, curve={height=6pt}, from=2-3, to=4-3]
	\arrow[color={rgb,255:red,153;green,92;blue,214}, curve={height=6pt}, from=1-4, to=3-4]
	\arrow["{{\small\exp(2\pi \ii\eta)}}", color={rgb,255:red,153;green,92;blue,214}, curve={height=6pt}, from=1-2, to=3-2]
\end{tikzcd}
\end{equation}
Notice that all groups in the lower triangle are isomorphic to $\Z/9$, and that since $j^{*}\circ \pi^{*} = \text{id}$, both these maps must be group isomorphisms.
Thus, if we know the anomalies for representations of $\Z/3$, and we know how to map representations of $A_4$ to and from representations of $\Z/3$, then commutativity of (\ref{eq:A4-4d-interplay}) allows us to deduce the anomaly for every irrep of $A_4$.\footnote{We emphasise that, in order to unambiguously relate $A_4$ anomalies to $\Z/3$ anomalies in this way, it is crucial that there are maps $j^\ast$ and $\pi^\ast$ going in both directions, which follows from the right-splitting of (\ref{eq:A4-SES}). (And this property is not even enough in general, if the bordism groups differ -- see \S \ref{sec:SL23} for the relevant example of $\SL(2,\F_3)=Q_8 \rtimes \Z/3$.)}

We first recap the anomalies for $\Z/3$. In general, all irreps of the cyclic group $\Z/n$ are one-dimensional. There is an irrep taking $1 \mod n \in \Z/n$ to $\exp(2\pi\ii q/n)$ for each $q=0,1,\ldots, n-1$, denoted here by $[q]_n$. In particular, for $\Z/3$,
we have three irreps, namely the trivial
representation $[0]_3$, and the fundamental representation $[1]_3$
together with its complex conjugate $[2]_3$. The exponentiated $\eta$-invariant maps, along the middle vertical arrow in (\ref{eq:A4-4d-interplay}), the fundamental representation $[1]_3$ to $1 \mod 9$ in $\mho^6_{\Spin}(B\Z/3)$~\cite{Hsieh:2018ifc}. 

We now relate the representations of $A_4$ and those of $\Z/3$ under
$j^{*}$ and $\pi^{*}$. Recall that $j^{*}$ maps an $A_4$ representation
to a $\Z/3$ representation. From the character table, it is clear that
\begin{equation}
  \begin{split}
    j^{*} : RU(A_4) & \to RU(\Z/3):\\
    \textbf{1} & \mapsto [0]_3\, ,\\
    \textbf{1}^{\prime} & \mapsto [1]_3\, ,\\
    \overline{\textbf{1}^{'}} & \mapsto [2]_3\, .
  \end{split}
\end{equation}
For the three-dimensional representation $\textbf{3}$, there is a basis where the order-3 generator
$\sigma$ of a $\Z/3$ subgroup acts by the $3 \times 3$ traceless matrix $\sigma = \diag(1,\omega, \omega^2)$,
where $\omega=\exp(2\pi\ii/3)$). Therefore,
\begin{equation}
j^{*}: \textbf{3} \to [0]_3 \oplus [1]_3 \oplus [2]_3
\end{equation}
for the triplet representation.

By the anomaly interplay diagram (\ref{eq:A4-4d-interplay}), we can then conclude that the trivial
representation $\textbf{1}$ and the three-dimensional irrep $\textbf{3}$ of $A_4$ are anomaly free, while the irreps $\textbf{1}^{\prime}$ and
$\overline{\textbf{1}^{\prime}}$ have non-trivial anomalies of order
$9$ with the opposite sign. Our result is consistent with the fact that chiral fermions in real representations cannot contribute to any anomaly in 4d \cite{Witten:2015aba}. (The triplet of $A_4$ is a real representation, which can be verified directly from the character table by applying the Frobenius--Schur indicator~(\ref{eq:FS}).)
The condition for $A_4$ anomaly cancellation in 4d is thus
\begin{equation}
\label{eq:A4-4d-anom-cancellation}
\boxed{n(\textbf{1}^{\prime})-n(\overline{\textbf{1}^{\prime}}) = 0 \mod 9\,  \qquad \text{($A_4$ anomaly cancellation in 4d),}}
\end{equation}
where $n(\textbf{r})$ denotes the number of left-handed fermions minus the number of right-handed fermions in the representation $\textbf{r}$.
There is no constraint on the number of (anomaly-free) triplet representations.
This information can be restated in a more succinct manner in terms of the subgroup of non-anomalous representations $RU_0(A_4)$ of $RU(A_{4})$. Our result shows that this group  is given by 
  \begin{equation}
    \label{eq:RU0A4}
RU_0(A_4) = \Z \left\langle \textbf{1} \right\rangle \oplus \Z \left\langle \textbf{1}^{\prime}\oplus \overline{\textbf{1}^{\prime}}\right\rangle \oplus \Z \left\langle \textbf{1}^{\oplus 9}  \right\rangle \oplus \Z \left\langle \textbf{3} \right\rangle.
\end{equation}

The full anomaly discussion above also enables us to relate to the notion of 
anomaly-free subgroups used in Refs. \cite{Kobayashi:2021xfs, Gripaios:2022vvc}. Notice that
there is an order $9$ anomaly for $A_4$ if and only if the fermions are
in either the $\textbf{1}^{'}$ representation or its complex
conjugate, which do not transform under the normal subgroup
$\Z/2 \times \Z/2$. On the other hand, the irrep
$\textbf{3}$ that transforms non-trivially under $\Z/2 \times \Z/2$ has no
anomaly. Thus, the anomaly-free subgroup of the alternating group
$A_{4}$ is its normal subgroup $\Z/2 \times \Z/2$, in agreement with the
result of \cite{Gripaios:2022vvc}.

In \S \ref{sec:A4trunc} we will tackle the issue of group extensions. We will there learn that the order 9 anomaly (\ref{eq:A4-4d-anom-cancellation}) can in fact be truncated down to an order 3 anomaly by a group extension of $A_4$ by $\Z/3$.

\subsection{$G=Q_8$ }
\label{sec:Q8-4d}

Next, we turn to the quaternion group $G=Q_8$, which has been used for example in quark mass model building~\cite{Frampton:1995fta}.
It is a multiplicative group of order $8$ whose elements consist of
$1$, and the unit quaternions $i,j,k$ ($i^2=j^2=k^2=-1$, $ij=k$),
together with their additive inverses. There are three non-trivial
1-dimensional irreps, $\textbf{1}_i$, $\textbf{1}_j$, $\textbf{1}_k$,
which are the sign representations with kernels $\{\pm 1, \pm i\}$,
$\{\pm 1,\pm j\}$, $\{\pm 1, \pm k\}$, respectively.  The only other non-trivial irrep is 2-dimensional, denoted $\textbf{2}$. The character table for $Q_8$ is given in Table \ref{tab:Q8-char}. 
\begin{table}[h]
  \centering
  \begin{tabular}{c|ccccc}
    & $1$ & $1$ & $2$ & $2$ & $2$\\
    $Q_8$ & $\{1\}$ & $\{-1\}$ & $\{i,-i\}$ & $\{j,-j\}$ & $\{k,-k\}$\\
    \hline
    $\textbf{1}$ & $1$ & $1$ &$1$ & $1$ & $1$\\
    $\textbf{1}_i$ & $1$ & $1$ & $1$ & $-1$ & $-1$\\
    $\textbf{1}_j$ & $1$ & $1$ & $-1$ & $1$ & $-1$\\
    $\textbf{1}_k$ & $1$ & $1$ & $-1$ & $-1$ & $1$\\
    $\textbf{2}$ & $2$ & $-2$ & $0$ & $0$ & $0$ 
  \end{tabular}
  \caption{Character table of $Q_8$, the quaternion group. The column are labelled by the conjugacy classes with the numbers of elements in each class atop.}
  \label{tab:Q8-char}
\end{table}
 
From the bordism perspective, this example appears to be rather interesting; the derived subgroup is $Q_8^\prime \cong \Z/2$ and the abelianisation is $Q_8/Q_8^\prime \cong \Z/2 \times \Z/2$, both of which are anomaly-free. But $G=Q_8$ is not anomaly-free, with 
\begin{equation}
\rbord_5(BQ_8) = \Z/2 \times \Z/2.
\end{equation} 
Thus, the anomalies for $Q_8$, and more generally for any finite group $G$, cannot be `pieced together' from the anomalies in its abelianisation and in its derived subgroup (with which one can `build' $G$ via the SES $G^\prime \to G \to G/G^\prime$), considered separately.

By applying the Frobenius--Schur indicator \eqref{eq:FS} to each of the irreps, we learn that all 1-dimensional irreps of $Q_8$ are real, while the 2-dimensional irrep is pseudo-real. Thus, the singlets are anomaly-free, but there could be an order 2 anomaly from fermions in the doublet representation. It turns out, however, that the $\eta$-invariant evaluated on generators of $\Omega^{\Spin}_5(BQ_8)$ vanishes \cite{Botvinnik:1995a}. Thus, 
\begin{equation}
\text{
All $Q_8$ representations are anomaly-free in 4d.}
\end{equation}
In this case, $RU_0(Q_8)$ is the whole of $RU(Q_8)$.

We remark that this is consistent with the anomaly interplay with $\Z/4$, using a subgroup embedding $\pi:\Z/4 \to Q_8$ (which is the only possibly-non-trivial anomaly interplay diagram involving a subgroup or quotient of $Q_8$). Under the pullback to the representation ring, one can check that every irrep of $Q_8$ pulls back to an anomaly-free representation of $\Z/4$. 

There nonetheless remains a curiosity with this $Q_8$ example, which is that the bordism group classifying anomalies is not zero, but $\Z/2\times \Z/2$. Even though there are no anomalies, the $Q_8$ structure allows non-trivial TQFT couplings that detect these bordism classes. We offer a partial account of these TQFTs in Appendix~\ref{app:tqft-Q8}, where we show that one of the bordism invariants is
\begin{equation}
\text{Arf}\cup xy^2,
\end{equation}
where $x,y$ are degree 1 generators of $H^{\bullet}(BQ_8;\Z/2)$ and $\text{Arf}$ is the Arf invariant of the spin structure.
This phenomenon that free fermions do not account for every possible anomaly given by the cobordism group has also been observed in 2d when the symmetry group is the dihedral group $D_8$ \cite{Grigoletto:2021oho}.

\subsection{$G=\SL(2,\mathbb{F}_3)$ } \label{sec:SL23}

We now consider the group $G=\SL(2,\mathbb{F}_3)$ of special linear matrices with elements in the finite field $\F_3$, that was used in flavour physics model building in Refs.~\cite{Carr:2007qw,Feruglio:2007uu}. This group has order-24, with 7 conjugacy classes of element and thus 7 irreps, with the character table shown in Table \ref{tab:SL2F-char}.
\begin{table}[h]
  \centering
  \begin{tabular}{c|ccccccc}
    & $1$ & $1$ & $4$ & $4$ & $6$ & $4$ & $4$\\
    $\SL(2,\mathbb{F}_3)$ & $\{\begin{psmallmatrix}1&0\\0&1\end{psmallmatrix}\}$ & $\{\begin{psmallmatrix}-1&0\\0&-1\end{psmallmatrix}\}$ & $\{\begin{psmallmatrix}1&-1\\0&1\end{psmallmatrix}\}$ & $\{\begin{psmallmatrix}1&1\\0&1\end{psmallmatrix}\}$ & $\{\begin{psmallmatrix}0&-1\\1&0\end{psmallmatrix}\}$ & $\{\begin{psmallmatrix}-1&1\\0&-1\end{psmallmatrix}\}$ & $\{\begin{psmallmatrix}-1&-1\\0&-1\end{psmallmatrix}\}$\\
    \hline
    $\textbf{1}$ & $1$ & $1$ &$1$ & $1$ & $1$ & $1$ & $1$\\
    $\textbf{1}^\prime$ & $1$ & $1$ &$\omega$ & $\omega^2$ & $1$ & $\omega$ & $\omega^2$\\
    $\overline{\textbf{1}^\prime}$ & $1$ & $1$ &$\omega^2$ & $\omega$ & $1$ & $\omega^2$ & $\omega$\\
    $\textbf{2}$ & $2$ & $-2$ &$-1$ & $-1$ & $0$ & $1$ & $1$\\
    $\textbf{2}^\prime$ & $2$ & $-2$ &$-\omega$ & $-\omega^2$ & $0$ & $\omega$ & $\omega^2$\\
    $\overline{\textbf{2}^\prime}$ & $2$ & $-2$ &$-\omega^2$ & $-\omega$ & $0$ & $\omega^2$ & $\omega$\\
    $\textbf{3}$ & $3$ & $3$ &$0$ & $0$ & $-1$ & $0$ & $0$\\
  \end{tabular}
  \caption{Character table for the group $\SL(2,\mathbb{F}_3)$. The columns are labelled by representatives of conjugacy classes, with the numbers of elements in each class written above.}
  \label{tab:SL2F-char}
\end{table}

By computing the Frobenius--Schur indicator (\ref{eq:FS}), we immediately learn that the irreps $\textbf{1}$, and $\textbf{3}$ are real and hence anomaly-free, while the irrep $\textbf{2}$ is pseudo-real and suffers at most an order 2 anomaly. The other four irreps (two singlets, and two doublets) are, however, complex. In this Section we analyse their anomalies using cobordism and anomaly interplay. This will be the most elaborate example we consider, in which a variety of techniques are needed to eventually arrive at the complete result~(\ref{eq:SL2F-4d-anom-cancellation}).

\subsubsection{Anomaly interplay with $\Z/3$ abelianisation}

Using the bordism calculation in Appendix~\ref{app:SL2F3}, we have that anomalies are here classified by the cobordism group
\begin{equation}
\mho^6_\Spin(B\SL(2,\mathbb{F}_3)) \cong \Omega_5^\Spin(B\SL(2,\mathbb{F}_3)) \cong \Z/2 \times \Z/2 \times \Z/9\, .
\end{equation}
\sloppy The derived subgroup of $\SL(2,\mathbb{F}_3)$ is the quaternion group $Q_8$, with abelianisation $\SL(2,\mathbb{F}_3)/Q_8 \cong \Z/3$. Indeed, the short exact sequence\footnote{We remark that there is also a short exact sequence $\Z/2 \hookrightarrow \SL(2,\F_3) \twoheadrightarrow A_4$, which follows from the fact that $\SL(2,\F_3)$ is a double cover of $A_4$. This sequence is less useful for our purposes because, unlike the abelianisation sequence (\ref{eq:abelianisation-SL23}), it is not right-split.
}
\begin{equation} \label{eq:abelianisation-SL23}
0 \to Q_8 \xrightarrow{i} \SL(2,\mathbb{F}_3) \xrightarrow{\pi} \Z/3 \to 0
\end{equation}
is right-split, meaning that $\SL(2,\mathbb{F}_3)$ has a semi-direct product structure
\begin{equation}
\SL(2,\mathbb{F}_3) = Q_8 \rtimes \Z/3,
\end{equation}
and that the abelianisation $\Z/3$ is also a subgroup, with subgroup embedding $j:\Z/3 \to \SL(2,\mathbb{F}_3)$. Recall from our previous examples that 4d anomalies for $G=\Z/3$ are classified by the cobordism group $\mho^6_\text{Spin}(B\Z/3) \cong \Z/9$, and that $\mho^6_\text{Spin}(BQ_8) \cong \Z/2\times \Z/2$. 

We will use these maps to study anomalies in the $\SL(2,\mathbb{F}_3)$ symmetry by anomaly interplay.
As usual, by applying the representation ring and cobordism functors to these maps, and relating the two via the exponentiated $\eta$-invariant, we obtain the following anomaly interplay commutative diagram\footnote{The homological version of this anomaly interplay diagram was a crucial ingredient in our calculation of the bordism group $\Omega_5^\Spin(B\SL(2,\mathbb{F}_3))$, as recorded in Appendix~\ref{app:SL2F3}. }
\begin{equation}\label{eq:SL2F-4d-interplay}
\begin{tikzcd}
	{RU(Q_8)} & {RU(\mathrm{SL}(2,\mathbb{F}_3))} && {RU(\Z/3)} \\
	{} & {} & {RU(\Z/3)} & {} \\
	{\mho_\mathrm{Spin}^6(BQ_8)} & {\mho_\mathrm{Spin}^6(B\mathrm{SL}(2,\mathbb{F}_3))} & {} & {\mho_\mathrm{Spin}^6(B\Z/3)} \\
	{} & {} & {\mho_\mathrm{Spin}^6(B\Z/3)}
	\arrow["{\qquad \pi^\ast}"', from=3-4, to=3-2]
	\arrow["{i^\ast}"', from=3-2, to=3-1]
	\arrow["{\exp(2\pi \ii \eta)}", color={rgb,255:red,153;green,92;blue,214}, curve={height=6pt}, from=1-1, to=3-1]
	\arrow["{j^\ast}", from=3-2, to=4-3]
	\arrow[Rightarrow, no head, from=3-4, to=4-3]
	\arrow["{j^\ast}", from=1-2, to=2-3]
	\arrow["{\pi^\ast}"', from=1-4, to=1-2]
	\arrow[Rightarrow, no head, from=1-4, to=2-3]
	\arrow["{i^\ast}"', from=1-2, to=1-1]
	\arrow[color={rgb,255:red,153;green,92;blue,214}, curve={height=6pt}, from=1-2, to=3-2]
	\arrow[color={rgb,255:red,153;green,92;blue,214}, curve={height=6pt}, from=2-3, to=4-3]
	\arrow[color={rgb,255:red,153;green,92;blue,214}, curve={height=6pt}, from=1-4, to=3-4]
\end{tikzcd}
\end{equation}
As groups, the entries in the `lower plane' are
\begin{equation}
\begin{tikzcd}
	{(\Z/2)^2} & {(\Z/2)^2\times \Z/9} & {} & {\Z/9} \\
	&& {\Z/9}
	\arrow["{j^\ast}", from=1-2, to=2-3]
	\arrow["{\pi^\ast}"', from=1-4, to=1-2]
	\arrow["{i^\ast}"', from=1-2, to=1-1]
	\arrow[Rightarrow, no head, from=1-4, to=2-3]
\end{tikzcd}
\end{equation}
This diagram should be contrasted with the equivalent  diagram (\ref{eq:A4-4d-interplay}) for the case $G=A_4$. In that case, the presence of maps $\pi^\ast$ and $j^\ast$ going in both directions, and the fact that the cobordism groups in the triangle were all isomorphic, guaranteed that $\pi^\ast$ and $j^\ast$ were isomorphisms. This meant that the anomalies for each irrep of $A_4$ could be deduced directly by decomposing to irreps of $\Z/3$, and computing the anomalies thereof.
Here, the extra factor of $(\Z/2)^2$ in the cobordism group for $\SL(2,\F_3)$ prevents such a direct method, and we will need extra ingredients to complete our analysis.

Proceeding, we first need to know the maps between representation rings in the upper plane. We can restrict our attention to the four complex irreps, and the one pseudo-real irrep, since only these can have anomalies.
The pullback of the complex representations of $\SL(2,\mathbb{F}_3)$ is
\begin{equation}
j^\ast: 
\textbf{1}^\prime \mapsto [1]_3, \quad\overline{\textbf{1}^\prime} \mapsto [2]_3, \quad  \textbf{2}^\prime \mapsto [0]_3 \oplus [2]_3,  \quad \overline{\textbf{2}^\prime} \mapsto [0]_3 \oplus  [1]_3\, 
\end{equation}
under the subgroup embedding $j$. On the other hand, the pullback of complex representations of $\Z/3$ under the quotient map $\pi$ is
\begin{equation} \label{eq:pi_star}
\pi^{*} : [1]_3 \mapsto \textbf{1}^{\prime}, \qquad [2]_3 \mapsto \overline{\textbf{1}^{\prime}}.
\end{equation}
Recall that the $[0]_3$ of $\Z/3$ is anomaly-free, while the $[1]_3$ and $[2]_3$ contribute $1 \mod 9$ and $-1 \mod 9$ anomalies respectively.

While we observed above that the maps $j^\ast$ and $\pi^\ast$ are not isomorphisms, as they were in the $G=A_4$ example of \S \ref{sec:A4-4d}, the fact that $j^\ast \circ \pi^\ast$ is the identity map is still extremely useful; it uniquely fixes the order 9 part of the anomaly for every irrep of $\SL(2,\mathbb{F}_3)$. Thence, using the maps between representations above, we can swiftly extract the following necessary condition for anomaly cancellation:
\begin{equation}
\label{eq:SL2F-4d-anom-cancellation}
n(\textbf{1}^{\prime})-n(\overline{\textbf{1}^{\prime}}) - n(\textbf{2}^{\prime})+n(\overline{\textbf{2}^{\prime}}) = 0 \mod 9\,.  
\end{equation}
Moreover, we can deduce that the anomalies of the irreps $\textbf{1}^{\prime}$ and $\overline{\textbf{1}^{\prime}}$, which (by (\ref{eq:pi_star})) are the only non-trivial irreps in the image of $\pi^\ast$, contain no $\Z/2 \times \Z/2$ element because $\pi^{*}$ is a homomorphism. Thus, so far we have computed the anomaly for all the 1-dimensional and 3-dimensional irreps of $\SL(2,\mathbb{F}_3)$. 

Unlike the $G=A_4$ example, however, this analysis does not yet fix
the anomaly completely; we must determine the $\Z/2 \times \Z/2$ part
of the anomaly for all three of the 2-dimensional irreps. We need a
new strategy to finish the task. It will turn out that the $\Z/2\times
\Z/2$ factor of $\Omega^{\Spin}_5(B\SL(2,\mathbb{F}_3))$ is not probed
by the $\eta$-invariant, just like the case for the quaternion group
$Q_8$ considered in Section \ref{sec:Q8-4d}.

\subsubsection{Anomaly interplay with $Q_8$ derived subgroup}

To see why this is the case, we will first argue that $\exp(2\pi\ii \eta)_{\SL(2,\mathbb{F}_3)}(\textbf{2})$ is trivial on any generator of the $\Z/2 \times \Z/2$ subgroup of $\Omega^{\Spin}_5(B\SL(2,\mathbb{F}_3))$, and so, by the fact that the irrep \textbf{2} is pseudo-real and can only have order 2 anomalies, it is anomaly-free.
To proceed, let $(X,f)$ denote a spin 5-manifold $X$ equipped with a certain $\SL(2,\mathbb{F}_3)$-bundle $f:X\to B\SL(2,\mathbb{F}_3)$, that is a generator of $\Z/2 \times \Z/2\subset \Omega^{\Spin}_5(B\SL(2,\mathbb{F}_3))$. It turns out that $(X,f')$, where $f':X\to BQ_8$ is naturally induced from $f$, is also a generator of $\Omega^{\Spin}_5(BQ_8)\cong \Z/2 \times \Z/2$, as we show in Appendix \ref{app:bord-gen-SL23}. By the pullback $i^{*}:\mho^6_{\Spin}(B\SL(2,\mathbb{F}_3))\to \mho^6_{\Spin}(BQ_8)$, we thus infer that
\begin{equation}
\exp \left( 2\pi\ii \eta\right)_{\SL(2,\mathbb{F}_3)}(\textbf{2},X) = \exp\left( 2\pi\ii \eta\right)_{Q_8}(\textbf{2},X),
\end{equation}
where we used the character table (Table~\ref{tab:SL2F-char}) to deduce that the pullback of the irrep $\textbf{2}$ of $\SL(2,\mathbb{F}_3)$ under the embedding $i$ is the irrep $\textbf{2}$ of $Q_8$. The RHS of this equation is trivial due to a result by Botvinnik and Gilkey  \cite{Botvinnik:1995a}, as used in \S\ref{sec:Q8-4d}.

\subsubsection{Anomaly interplay with $\UU(2)$: proof `by Standard Model'}
\label{sec:interplay-with-SM}

All that remains is to finish computing the anomaly for the complex 2-dimensional irreps.
To show that these do not contribute any order 2 anomaly, we now study an anomaly interplay between $\SL(2,\mathbb{F}_3)$ and $\UU(2)$. Recall that the group $\SL(2,\mathbb{F}_3)$ is a semi-direct product between the quaternion group $Q_8$ and $\Z/3$ where  $\Z/3$ acts on $Q_8$ by cyclically permuting the quaternionic units $i,j,k$. More precisely, we can write
\begin{equation}
\SL(2,\mathbb{F}_3) = \left\langle i,j,\sigma| i^4=\sigma^3=1, j^2=i^2, jij^{-1}=i^{_1},\sigma i \sigma^{-1}=j, \sigma j \sigma^{-1}=ij \right\rangle\, .
\end{equation}
Using this, we now find an explicit embedding of $\SL(2,\mathbb{F}_3)$ as a subgroup of $\UU(2)$, defining a group homomorphism $s:\SL(2,\mathbb{F}_3)\to \UU(2)$ such that
\begin{equation}\label{eq:SL23-U2-embedding}
  s:  i \mapsto
    \begin{pmatrix}
      \ii & 0\\ 0 & -\ii
    \end{pmatrix},\qquad 
    j \mapsto
    \begin{pmatrix}
      0 & -1\\ 1 & 0
    \end{pmatrix},\qquad
    \sigma \mapsto
    \ee^{\pi \ii /12}\begin{pmatrix}
      1 & -1\\ -\ii & -\ii
    \end{pmatrix}.
\end{equation}
It is straightforward to verify that the relations determining the structure of $\SL(2,\mathbb{F}_3)$ are all satisfied with this choice of embedding. 

We do not need the full machinery of anomaly interplay between local and global anomalies used in Ref. \cite{Davighi:2020uab} because we only consider $\UU(2)$ representations that are free of local anomalies. The only thing we will make use of here is that anomaly-free representations remain anomaly-free under the pullback. Let us denote an irrep of $\UU(2) = \left( \SU(2) \times \U \right)/(\Z/2)$ by $(\textbf{n},q)$, where $\textbf{n}$ is the irrep under the $\SU(2)$ subgroup and $q$ is the $\U$ charge satisfying $q-(n-1)\in 2\Z$.

Consider Weyl fermions in a well-known anomaly-free representation of $\UU(2)$, namely one generation of the Standard Model fermion content whose representation under the electroweak gauge group $\UU(2)$ reads\footnote{Here, local physics allows us to choose the global form of the gauge group to be $\SU(3) \times \UU(2)$. For other possibilities of the gauge group for the algebra $\g = \su(3)\oplus \su(2)\oplus \u(1)$ and ways to distinguish between them, see {\em e.g.} Ref. \cite{Tong:2017oea}.}
\begin{equation}
  \label{eq:SM-rep}
\textbf{r}_{\text{SM}} =  \left( \textbf{2},1 \right)^{\oplus 3} \oplus \left( \textbf{2},-3 \right) \oplus  \left( \textbf{1},-4 \right)^{\oplus 3} \oplus  \left( \textbf{1},2 \right)^{\oplus 3} \oplus \left( \textbf{1},6 \right) .
\end{equation}
Restricting to the subgroup $\SL(2,\mathbb{F}_3)$ under the embedding $s$ given in \eqref{eq:SL23-U2-embedding}, the representation $\textbf{r}_{\text{SM}}$ decomposes in terms of the irreps of $\SL(2,\mathbb{F}_3)$ as (see Appendix~\ref{app:SM})
\begin{equation}
  \label{eq:SM-decomp}
s^{*}\textbf{r}_{\text{SM}}=  \textbf{2}^{\prime\,\oplus 3} \oplus \textbf{2} \oplus   \overline{\textbf{1}^{\prime}}^{\,\oplus 6}  \oplus \textbf{1} \,.
\end{equation}
Note that in this representation, $n(\textbf{1}^{\prime})-n(\overline{\textbf{1}^{\prime}}) - n(\textbf{2}^{\prime})+n(\overline{\textbf{2}^{\prime}})=-9$, consistent with the anomaly cancellation condition \eqref{eq:SL2F-4d-anom-cancellation}.

Now, let's evaluate the $\eta$-invariant in this representation on a generator $X$ of $\Z/2\times \Z/2\subset\Omega^{\Spin}_5(B\SL(2,\mathbb{F}_3))$ (which is also a manifold equipped with a $\UU(2)$-bundle). Since we know that this representation is anomaly-free, we get
\begin{equation}
1= \exp\left(2\pi\ii \eta\right)_{\SL(2,\mathbb{F}_3)}(s^{*}\textbf{r}_{\text{SM}}, X) = \exp\left(2\pi\ii \eta\right)_{\SL(2,\mathbb{F}_3)}(\textbf{2}^{\prime\,\oplus 3},X),
\end{equation}
where we use the fact that all other irreps in $s^{*}\textbf{r}_{\text{SM}}$ has no order 2 anomaly in the last equality. This immediately tells us that the irrep $\textbf{2}^{\prime}$ and its conjugate are free of any  anomaly of order 2.

Putting everything together, we have proven that all representations of $\SL(2,\mathbb{F}_3)$ have no order 2 anomaly, and moreover computed the anomalies associated with every irrep. Thus, the condition 
\begin{equation}
  \tag{\ref{eq:SL2F-4d-anom-cancellation}}
\boxed{n(\textbf{1}^{\prime})-n(\overline{\textbf{1}^{\prime}}) - n(\textbf{2}^{\prime})+n(\overline{\textbf{2}^{\prime}}) = 0 \mod 9\,  \qquad \text{($\SL(2,\mathbb{F}_3)$ anomaly cancellation in 4d)}}
\end{equation}
is both necessary and sufficient for $\SL(2,\mathbb{F}_3)$ anomaly cancellation. The anomaly cancellation condition above is equivalent to saying that the non-anomalous representations are generated by the set
  \begin{equation}
    \label{eq:RU0SL}
\textbf{1}, \quad \textbf{2}, \quad \textbf{3}, \quad \textbf{1}^{\prime}\oplus \overline{\textbf{1}^{\prime}}, \quad \textbf{2}^{\prime}\oplus \overline{\textbf{2}^{\prime}}, \quad \textbf{1}^{\prime} \oplus \overline{\textbf{2}^{\prime}}, \quad\textbf{1}^{\prime\,\oplus 9} \,,
\end{equation}
which can be taken as a basis for  $RU_0(\SL(2,\mathbb{F}_3))$.

\subsubsection{Example: an anomalous flavour physics model} \label{sec:anomalous-flavour}

One can apply our result to compute the anomaly in various flavour physics models that use an $\SL(2,\F_3)$ symmetry. For an explicit example, we consider the model of Feruglio et al.~\cite{Feruglio:2007uu} for quark and lepton masses. We denote the fermion representation there used, recorded in Table 2 of Ref.~\cite{Feruglio:2007uu}, by ${\bf r}_{\mathrm{BSM}}$.
The leptons are charged in triplet and singlet representations of $\SL(2,\F_3)$, to mimic the construction of $A_4$-based models for lepton masses and mixings. The doublet representations are also utilised, to unify the first two generations of quarks, while the third family quarks transform in singlets.

Within the leptons alone, one can check that anomalies cancel. But summing over the quarks, there are 12 left-handed Weyl fermions transforming in one of the complex doublet irreps, which we can take to be the irrep ${\bf \overline{2^\prime}}$ without loss of generality. From the results of this Section, the anomaly is therefore
\begin{equation}
\exp(2\pi \ii \eta)_{\SL(2,\F_3)}({\bf r}_{\mathrm{BSM}}) =\exp(2\pi \ii/3) \, .
\end{equation}
Interestingly, the anomaly is non-zero, corresponding to the element $3 \mod 9 \in \mho_6^\Spin(B\SL(2,\F_3))$. The $\SL(2,\F_3)$ symmetry is therefore anomalous. Perhaps more interestingly, this is not the end of the story, as we shall see in \S \ref{sec:flavour-truncation}; it turns out this anomaly can be `truncated' (in a sense soon to be defined) by passing to a larger symmetry group that surjects onto $\SL(2,\F_3)$.

\subsection{Dihedral group symmetry}

Our last example is the symmetric group $S_3$, whose anomaly is also analysed in Ref.~\cite{Gripaios:2022vvc} by a different method. From the bordism perspective this example is especially simple, because the fact that
\begin{equation}
\Omega_5^\Spin(BS_3) = 0
\end{equation}
means there can be no anomalies in 4d. To see this, recall that $S_3$ is isomorphic to the dihedral group of order $6$, $D_6$. Further, as we explain in Appendix \ref{app:D2n}, it can be shown that  
\begin{equation}
\Omega_5^\Spin(BD_{2n}) = 0
\end{equation}
for any integer $n$, so any such dihedral group symmetry (of which $S_3$ is a special case) is anomaly-free in 4d.

\section{Not all anomalies are equal: anomaly-truncation by group extensions}
\label{sec:truncation}

In the previous Section we computed the anomalies associated with all representations for certain non-abelian finite groups.
But there is a subtle issue that we must confront when considering anomalies for finite groups. This is the issue of whether a finite group anomaly can be `truncated' (by which we mean, very roughly, that the order of the (finite) anomaly for a given representation is reduced) by an extension of the symmetry.
This idea was first appreciated, for the case of abelian finite groups, by Banks and Dine~\cite{Banks:1991xj} (and discussed from the bordism perspective in~\cite{Hsieh:2018ifc}).

In \S \ref{sec:trunc-formalism} we put this notion of `anomaly truncation' on a rigorous footing. We then review the known example of Banks and Dine in \S \ref{sec:Banks-Dine}, recasting the analysis in our language, before considering examples with non-abelian finite symmetry groups in \S \S \ref{sec:A4trunc} and \ref{sec:SL23-trunc}.

\subsection{General formalism}
\label{sec:trunc-formalism}

Consider a group extension $f:H\twoheadrightarrow G$ of $G$. Since this is a surjection, for any element of $G$ there is a corresponding element in $H$, and so we retain the original symmetry of the theory. One could then ask about the anomaly of the system of fermions considered as transforming in a representation of $H$ instead of the original $G$.

As discussed in \S \ref{sec:cobord-class}, the homomorphism $f$ induces the commutative diagram 
\begin{equation} \label{eq:ROm_square-2}
\begin{tikzcd}
{RU(G)} & {RU(H)} \\
{\mho^6_{\text{Spin}}(BG)} & {\mho^6_{\text{Spin}}(BH)}
\arrow["{RU(f)}"', from=1-1, to=1-2]
\arrow["{\mho(f)}"', from=2-1, to=2-2]
\arrow["{\exp (2\pi \text{i}\eta)_G}"', from=1-1, to=2-1]
\arrow["{\exp (2\pi\text{i}\eta)_H}", from=1-2, to=2-2]
\end{tikzcd}\, .
\end{equation}
Clearly, an anomaly-free representation $\textbf{r}$ of $G$ remains anomaly-free after getting pulled back to a representation of $H$ since the commutativity implies that
  \begin{equation}
 \exp(2\pi\ii \eta)_{H} \left(RU(f) (\textbf{r}) \right) = \mho(f)\circ\exp (2\pi\ii \eta)_{G} (\textbf{r}) = 0.
  \end{equation}
It is also possible that this equation is satisfied for an anomalous representation $\textbf{r}$ of $G$. In other words, an anomalous representation may become anomaly-free after the group extension. In that case, we say that the anomaly is {\em truncated}, because the order of the image of $\exp(2\pi\ii \eta)_{H} \circ RU(f)$ is generally smaller than the order of the image of $\exp(2\pi\ii \eta)_{G}$.

This anomaly truncation can be quantified as follows. Recall that
\begin{equation}
RU_0(G) = \ker \left( \exp (2\pi\ii \eta)_G \right) \subset RU(G)
\end{equation}
is the subgroup of anomaly-free representations inside $RU(G)$. We
can similarly define the subgroup of representations of $G$ that
become anomaly-free after being pulled back along $f$ to be 
\begin{equation}
RU_0(G;H,f) := \ker \left( \mho(f)\circ \exp(2\pi\ii \eta)_{G}\right) \subset RU(G).
\end{equation}
Clearly, $RU_0(G)$ is a subgroup of $RU_0(G;H,f)$ since all anomaly-free representations are pulled back to anomaly-free representations. Thus, the anomaly truncation occurs when the quotient
\begin{equation}
Q(G;H, f) = RU_0(G;H,f)\big/RU_0(G)
\end{equation}
is non-trivial. We call $Q(G; H, f)$ the {\em anomaly truncation quotient}. 

\subsection{Known example: abelian finite anomalies } \label{sec:Banks-Dine}

To give this rather formal description some teeth, we start by revisiting the known example of anomaly truncation for a finite abelian group~\cite{Banks:1991xj,Hsieh:2018ifc}.

For concreteness, let us consider a 4d theory with symmetry $G=\Z/3$, that we would like to interpret as a low-energy effective field theory. As seen in \S \ref{sec:A4-4d}, the anomalies are captured by the cobordism group
$\mho^6_\Spin(B\Z/3) \cong \Omega_5^{\text{Spin}}(B\mathbb{Z}/3) \cong \Z/9$,
which we take to be generated by the exponentiated $\eta$-invariant for the fundamental $[1]_3$ representation.
For a generic set of LH Weyl fermions  the condition for $\Z/3$ anomaly cancellation is~\cite{Hsieh:2018ifc}.
\begin{equation} \label{eq:Z3-ACC}
n([1]_3)- n([2]_3)= 0 \mod 9\, .
\end{equation}
This can be derived, for example, by embedding $\Z/3 \hookrightarrow \U$. From that perspective, the mod 9 condition (\ref{eq:Z3-ACC}) comes from the cubic $\U$ anomaly cancellation condition (which implies the `coarser' linear condition, $n([1]_3)- n([2]_3) = 0 \mod 3$).

Now consider extending $\Z/3$ via
\begin{equation}
\begin{tikzcd}
	{\mathbb{Z}/3} & {\mathbb{Z}/9} & {\mathbb{Z}/3}
	\arrow["\alpha", hook, from=1-1, to=1-2]
	\arrow["p", two heads, from=1-2, to=1-3]
\end{tikzcd}\, .
\end{equation}
The pullback of representations is simply multiplication of charges by three, {\em viz.}
\begin{equation}
p^\ast: RU(\Z/3) \to RU(\Z/9) : [q]_3 \mapsto [3q]_9\, .
\end{equation}
Physically,
the key point is that this extension of the symmetry group from $\Z/3$ to $\Z/9$ is undetectable from the low-energy perspective. Indeed, there may lurk UV states with discrete charges that are finer multiples of the charges of the low-energy degrees of freedom (in this case {\em e.g.} $q_{UV}=[1]_9$), such that it is $\Z/9$ (or indeed any extension of $\Z/3$) that acts faithfully in the UV. 

For a general $\Z/9$ gauge theory, with a set of LH Weyls in the representation $\bigoplus_i [Q_i]_9$ of $\Z/9$, let us define the anomaly coefficients
$\mathcal{A}_{\text{cub}} :=\sum_i Q_i^3$ and $\mathcal{A}_{\text{grav}} :=\sum_i Q_i$ (so named because they pullback from the cubic and mixed-gravitational anomalies under embedding $\Z/9$ in $\U$).
The anomaly  associated to this fermion spectrum is~\cite{Hsieh:2018ifc}
\begin{equation}
\left(\mathcal{A}_{\text{cub}} \mod 27,\,\, \frac{1}{3}\left(\mathcal{A}_{\text{cub}}-\mathcal{A}_{\text{grav}}\right) \mod 3 \right) \in \mho^6_{\text{Spin}}(B\mathbb{Z}/9)\cong \Z/27 \times \Z/3 \, .
\end{equation}
A set of LH fermions in the representation $\bigoplus_i [q_i]_3$ of $\Z/3$ pulls back to $\bigoplus_i [3q_i]_9$ of $\Z/9$. Clearly, the $\Z/27$-valued element of the anomaly always vanishes after this pullback, since $(3q_i)^3 = 27 q_i^3 \implies \mathcal{A}_{\text{cub}} = 0 \mod 27$. This leaves only the order 3 anomaly, which vanishes if and only if $\sum_iq_i = 0 \mod 3$, which is equivalent to the condition
\begin{equation} \label{eq:Z3-trunc}
n([1]_3) - n([2]_3) = 0 \mod 3
\end{equation}
for fermions in the original $\Z/3$ theory. By comparison with the original condition (\ref{eq:Z3-ACC}), we see that the order of the anomaly is truncated from 9 to 3 by the symmetry extension.

We can state the anomaly truncation more clearly as the enlargement of the subgroup of anomaly-free representations. Originally, $RU_0(\Z/3)$ is generated by $[0]_3$, $[1]_3 \oplus [2]_3$, and $[1]_3^{\oplus 9}$. After the group extension to $\Z/9$,  the subgroup of anomaly-free representations becomes $RU_0(\Z/3)^{\prime}$, now generated by $[0]_3$, $[1]_3\oplus [2]_3$, and $[1]_3^{\oplus 3}$. Thus, the anomaly truncation quotient is
\begin{equation}
Q(\Z/3;\Z/9, p) = \frac{\Z \left\langle [1]_3^{\oplus 3} \right\rangle}{\Z \left\langle [1]_3^{\oplus 9} \right\rangle} \cong \Z/3.
\end{equation}

Note that the remaining condition (\ref{eq:Z3-trunc}) is itself the result of pulling back the linear mixed $\U$-gravitational anomaly along $\Z/9 \hookrightarrow \U$. Since that condition is {\em linear} in the charges, it is invariant under further symmetry extensions of $\Z/9$ by any $\Z/r$, which are tantamount simply to rescalings of all charges by $r$.

At least in this example, where $G$ is abelian, the anomaly truncation is equivalent to cancelling the anomaly by coupling to a topological quantum field theory (TQFT), as discussed {\em e.g.} in Ref.~\cite{Garcia-Etxebarria:2018ajm} (Section 4.6.2). The TQFT discussed therein has the effect of restricting the allowed $\Z/3$ bundles to those that lift to $\Z/9$ bundles. It will be interesting to see if this version of events, in which anomaly truncation via symmetry extension is explicitly realised by coupling to a TQFT, carries over to the case where $G$ is non-abelian, as in the examples we consider next. We will explore this more in future work.

\subsection{Extending  $A_4$ by  $\Z/3$} \label{sec:A4trunc}

Let us return to the $G=A_4$ anomalies in 4d, for which we have derived in \S \ref{sec:A4-4d} that the non-trivial singlet irreps $\textbf{1}^{\prime}$ and $\overline{\textbf{1}^{\prime}}$ contribute anomalies of $\pm 1$ mod $9$. The anomaly-free subgroup of $A_4$ was identified to be the derived subgroup $A_4^\prime \cong \Z/2 \times \Z/2$, with the anomaly `coming from' its abelianisation $A_4/A_4' \cong \Z/3$. 
Now, as we reviewed in \S \ref{sec:Banks-Dine} for the case of $\Z/3$ anomalies in 4d, the order 9 anomaly is truncated to an order 3 anomaly by extending the symmetry from $\Z/3$ to $p:\Z/9 \to \Z/3$. As discussed, such a symmetry extension is undetectable from the low-energy perspective, and so it is only the reduced mod 3 condition that {\em must} be satisfied by the $\Z/3$-charged IR degrees of freedom.

Since the order 9 gauge anomaly associated with the $\textbf{1}^{\prime}$ irrep of $G=A_4$ comes from its $\Z/3$ subgroup, in a precise sense captured by the anomaly interplay maps \eqref{eq:A4-4d-interplay}, it is plausible that this order 9 anomaly could also be truncated to, say, an order 3 anomaly by some finite group extension of $A_4$.
By analogy with the $G=\Z/3$ case above, we consider the extension 
\begin{equation}
0 \to \Z/3 \to K \to A_4 \to 0,
\end{equation}
of $A_4$ by $\Z/3$, where $K$ is the semi-direct product
\begin{equation}
K = (\Z/2 \times \Z/2) \rtimes \Z/9\, .
\end{equation}
\sloppy The injective map from $\Z/3$ takes the element $1 \mod 3 \in \Z/3$ to the element $((0 \mod 2, 0 \mod 2), 3 \mod 9)$ of $(\Z/2 \times \Z/2)\rtimes \Z/9$.
The resulting group $K$ can also be described as a different group extension, 
\begin{equation}
0 \to \Z/2 \times \Z/2 \to K \to  \Z/9 \to 0 \, ,
\end{equation}
which splits because $K$ is a semi-direct product.
Armed with both these extensions, we know that $K$ sits at the centre of the following commutative diagram of short exact sequences:
\begin{equation} \label{eq:grid}
\begin{tikzcd}
	0 & {\mathbb{Z}/2\times \mathbb{Z}/2} & {\mathbb{Z}/2\times \mathbb{Z}/2} \\
	{\mathbb{Z}/3} & K & {A_4} \\
	{\mathbb{Z}/3} & {\mathbb{Z}/9} & {\mathbb{Z}/3}
	\arrow[Rightarrow, no head, from=1-2, to=1-3]
	\arrow[from=1-1, to=1-2]
	\arrow[from=1-1, to=2-1]
	\arrow["{\pi^\prime}", from=2-2, to=3-2]
	\arrow["{i^\prime}", hook, from=1-2, to=2-2]
	\arrow["{j^\prime}", curve={height=-12pt}, dashed, from=3-2, to=2-2]
	\arrow[Rightarrow, no head, from=2-1, to=3-1]
	\arrow["\alpha"', hook, from=3-1, to=3-2]
	\arrow["{\alpha^\prime}", hook, from=2-1, to=2-2]
	\arrow["{p^\prime}", from=2-2, to=2-3]
	\arrow["i", hook, from=1-3, to=2-3]
	\arrow["\pi", from=2-3, to=3-3]
	\arrow["p"', from=3-2, to=3-3]
	\arrow["j", curve={height=-12pt}, dashed, from=3-3, to=2-3]
\end{tikzcd}\, .
\end{equation}
Importantly for the calculation that follows, the fact that both the middle and right columns are in fact split extensions means there are two maps `going the other way', labelled $j:\Z/3 \to A_4$ (a map we have already used extensively above) and $j^\prime:\Z/9\to K$, satisfying $j \circ \pi = \text{id}$ and $j^\prime \circ \pi^\prime=\text{id}$.\footnote{Note that the middle row, which describes $K$ as an extension of $A_4$, cannot split; if it did, that would imply a right-to-left map from $\Z/3$ to $\Z/9$ in the bottom row whose composition with $p$ is the identity, meaning that the bottom row splits also -- which it does not. }

For every one of the maps in the commutative diagram \eqref{eq:grid}, functorality implies an induced pullback on both the representation ring $RU(\cdot)$ and cobordism $\mho_\Spin^6(B\cdot)$, with commuting squares involving the exponentiated $\eta$-invariant that maps $RU(\cdot)\to\mho_\Spin^6(B\cdot)$, as drawn in \eqref{eq:ROm_square}.
For example, there is a commuting square:
\begin{equation}
\begin{tikzcd}
{RU(A_4)} & {RU(K)} \\
{\mho^6_{\text{Spin}}(BA_4)} & {\mho^6_{\text{Spin}}(BK)}
\arrow["{p^{\prime\ast}}"', from=1-1, to=1-2]
\arrow["{p^{\prime\ast}}"', from=2-1, to=2-2]
\arrow["{\exp (2\pi \text{i}\eta)_{A_4}}"', from=1-1, to=2-1]
\arrow["{\exp (2\pi\text{i}\eta)_K}", from=1-2, to=2-2]
\end{tikzcd}
\end{equation}
If we focus on the maps in the bottom right square of Eq.~\eqref{eq:grid}, this gives rise to the following commutative `cube',
\begin{equation} \label{eq:cube}
\begin{tikzcd}
	{} && {RU(K)} \\
	{RU(A_4)} &&&& {RU(\mathbb{Z}/9)} \\
	&& {RU(\mathbb{Z}/3)} && {} \\
	\\
	&& {\mho^6_{\text{Spin}}(BK)} & {} \\
	{\mho^6_{\text{Spin}}(BA_4)} &&&& {\mho^6_{\text{Spin}}(B\Z/9)} \\
	&& {\mho^6_{\text{Spin}}(B\mathbb{Z}/3)}
	\arrow["{p^{\prime\ast}}", from=2-1, to=1-3]
	\arrow["{j^{\prime\ast}}", from=1-3, to=2-5]
	\arrow["{j^\ast}"', from=2-1, to=3-3]
	\arrow["{p^\ast}"', from=3-3, to=2-5]
	\arrow["{p^{\prime\ast}}", from=6-1, to=5-3]
	\arrow["{j^{\prime\ast}}", from=5-3, to=6-5]
	\arrow["{j^\ast}"', from=6-1, to=7-3]
	\arrow["{p^\ast}"', from=7-3, to=6-5]
	\arrow["{\exp(2\pi \ii\eta)}", color={rgb,255:red,153;green,92;blue,214}, curve={height=18pt}, from=2-1, to=6-1]
	\arrow[color={rgb,255:red,153;green,92;blue,214}, curve={height=18pt}, from=3-3, to=7-3]
	\arrow[color={rgb,255:red,153;green,92;blue,214}, curve={height=18pt}, dashed, from=1-3, to=5-3]
	\arrow[color={rgb,255:red,153;green,92;blue,214}, curve={height=18pt}, dashed, from=2-5, to=6-5]
	\arrow["{\pi^{\prime\ast}}"', color={rgb,255:red,214;green,92;blue,92}, curve={height=-18pt}, from=6-5, to=5-3]
	\arrow["{\pi^\ast}", color={rgb,255:red,214;green,92;blue,92}, curve={height=-18pt}, from=7-3, to=6-1]
\end{tikzcd}
\end{equation}
where each `vertical face' is a commutative square of the kind \eqref{eq:ROm_square}.

Now, we wish to follow the fate of the representation ${\bf 1}^\prime \in RU(A_4)$, to see if its anomaly is truncated by passing to the extension $K$. Our computation of \S \ref{sec:A4-4d} means that
\begin{equation}
\exp(2\pi \ii \eta)_{A_4} : RU(A_4) \to \mho^6_{\text{Spin}}(BA_4): {\bf 1}^\prime \mapsto 1 \mod 9\, .
\end{equation}
We wish to compute $\exp(2\pi \ii \eta)_{K}(p^{\prime\ast}({\bf 1}^\prime)) \in \mho^6_{\text{Spin}}(BK)$, {\em i.e.} the anomaly associated with the pullback of the ${\bf 1}^\prime \in RU(A_4)$ to a $K$-representation. This equals $p^{\prime\ast}\left(\exp(2\pi \ii \eta)_{A_4}({\bf 1}^\prime)\right)$ by commutativity (of the `back left' face of \eqref{eq:cube}).
To find this anomaly pullback, we use our knowledge of the anomaly interplay maps $j^\ast$ and $p^\ast$ in the lower face of \eqref{eq:cube}. Recalling $j^\ast({\bf 1}^\prime) = [1]_3 \in RU(\Z/3)$,
we have
\begin{equation}
j^\ast: [\exp(2\pi \ii \eta)_{A_4}({\bf 1}^\prime) = \exp(2\pi \ii/9)] \mapsto [\exp(2\pi \ii \eta)_{\Z/3}([1]_3) =  \exp(2\pi \ii/9)]\, ,
\end{equation}
using the results of \S \ref{sec:A4-4d}. Now, using $p^\ast([1]_3) = [3]_9 \in RU(\Z/9)$, we have
\begin{equation}
p^\ast: [\exp(2\pi \ii \eta)_{\Z/3}([1]_3) =  \exp(2\pi \ii/9)] \mapsto [\exp(2\pi \ii \eta)_{\Z/9}([3]_9) =  \exp(2\pi \ii/3)]\, ,
\end{equation}
by the calculation in \S \ref{sec:Banks-Dine}.

Thus, if we restrict to the image of the anomalous representation ${\bf 1}^\prime \in RU(A_4)$ in the lower square, we have
\begin{equation}
\begin{tikzcd}
	&& {?} \\
	{1 \text{~mod~} 9} &&&& {1 \text{~mod~} 3} \\
	&& {1 \text{~mod~} 9}
	\arrow["{j^\ast}"', from=2-1, to=3-3]
	\arrow["{p^{\prime\ast}}", from=2-1, to=1-3]
	\arrow["{j^{\prime\ast}}", from=1-3, to=2-5]
	\arrow["{\pi^{\prime\ast}}"', color={rgb,255:red,214;green,92;blue,92}, curve={height=-18pt}, from=2-5, to=1-3]
	\arrow["{\pi^\ast}", color={rgb,255:red,214;green,92;blue,92}, curve={height=-18pt}, from=3-3, to=2-1]
	\arrow["{p^\ast}"', from=3-3, to=2-5]
\end{tikzcd}
\end{equation}
 where the question mark denotes $ \exp(2\pi \ii \eta)_K(p^{\prime\ast}({\bf 1}^\prime))$, the anomaly that we wish to pin down. Thanks to the existence of both maps $j^{\prime\ast}$ and $\pi^{\prime\ast}$ satisfying $j^{\prime\ast} \circ \pi^{\prime\ast} = \text{id}$ ({\em i.e.} thanks to the splitting of $\Z/2 \times \Z/2 \hookrightarrow K \twoheadrightarrow \Z/9$), we infer that the unknown anomaly phase is 
\begin{equation}
\exp(2\pi \ii \eta)_K(p^{\prime\ast}({\bf 1}^\prime)) =  \exp(2\pi \ii/3) \, . 
\end{equation}
Thus, by passing to the extension $K$, we see that the order 9 anomaly associated with the non-trivial singlets of $A_4$ can be truncated to an order 3 anomaly.

Alternatively, we could phrase this result in terms of the anomaly-free representations. After the group extension, the subgroup $RU_0(A_4)$ of anomaly-free representations becomes $RU_0(A_4; K, p^\prime)$, now generated by
\begin{equation}
\textbf{1},\quad \textbf{3},\quad \textbf{1}^{\prime}\oplus \overline{\textbf{1}^{\prime}}, \quad \textbf{1}^{\prime\,\oplus 3}.
\end{equation}
By taking the quotient with $RU_0(A_4)$ given in \eqref{eq:RU0A4}, it can be easily seen that the anomaly truncation quotient $Q(A_4; K, p^\prime)$ is isomorphic to $\Z/3$, showing that the order of the anomaly is `down' by a third.

We emphasise that, to calculate this anomaly, we did not even need to know the pullback representation $p^{\prime \ast}({\bf 1}^\prime) \in RU(K)$, nor did we need to know the cobordism group $\mho^6_{\text{Spin}}(BK)$. The `gridlock' properties of so many commutative squares, thanks to the two different short exact sequences for $K$, are powerful enough to completely fix the value of the anomaly that we wanted.

\subsection{Extending $\SL(2,\mathbb{F}_3)$ by $\Z/3$}
\label{sec:SL23-trunc}

Let's now consider the symmetry group $\SL(2,\mathbb{F}_3)$ and explore a possibility of anomaly truncation by group extension. First recall that, as shown in \S \ref{sec:SL23}, the only anomalies from Weyl fermions in representations of this symmetry group are order 9 anomalies, which directly comes from the order 9 anomalies associated with its $\Z/3$ subgroup. It is thus probable that these order 9 anomalies can be truncated in a similar manner to that of the symmetry group $A_4$ discussed in the previous Subsection.

We extend $\SL(2,\mathbb{F}_3)$ non-trivially by $\Z/3$ to the group
\begin{equation}
L = Q_8 \rtimes \Z/9
\end{equation}
that fits inside the commutative diagram
\begin{equation} \label{eq:SL23-grid}
\begin{tikzcd}
	0 & {Q_8} & {Q_8} \\
	{\mathbb{Z}/3} & L & {\SL(2,\mathbb{F}_3)} \\
	{\mathbb{Z}/3} & {\mathbb{Z}/9} & {\mathbb{Z}/3}
	\arrow[Rightarrow, no head, from=1-2, to=1-3]
	\arrow[from=1-1, to=1-2]
	\arrow[from=1-1, to=2-1]
	\arrow["{\pi^\prime}", two heads, from=2-2, to=3-2]
	\arrow["{i^\prime}", hook, from=1-2, to=2-2]
	\arrow["{j^\prime}", curve={height=-12pt}, dashed, from=3-2, to=2-2]
	\arrow[Rightarrow, no head, from=2-1, to=3-1]
	\arrow["\alpha"', hook, from=3-1, to=3-2]
	\arrow["{\alpha^\prime}", hook, from=2-1, to=2-2]
	\arrow["{p^\prime}", two heads, from=2-2, to=2-3]
	\arrow["i", hook, from=1-3, to=2-3]
	\arrow["\pi", two heads, from=2-3, to=3-3]
	\arrow["p"', two heads, from=3-2, to=3-3]
	\arrow["j", curve={height=-12pt}, dashed, from=3-3, to=2-3]
\end{tikzcd}\, ,
\end{equation}
where the injective map $\alpha^\prime$ maps $1 \mod 3$ in $\Z/3$ to $(0, 3 \mod 9)$ in $L$, and $\alpha$ is multiplication by 3.
Again, applying the functors $RU(\cdot)$ and $\mho^6_{\Spin}(\cdot)$ to the lower-right square of the grid \eqref{eq:SL23-grid}, we obtain two commutative squares linked together by the natural transformation $\exp(2\pi \ii \eta)$ to form a commutative cube 
\begin{equation} \label{eq:SL23-cube}
\begin{tikzcd}
	{} && {RU(L)} \\
	{RU(\SL(2,\mathbb{F}_3))} &&&& {RU(\mathbb{Z}/9)} \\
	&& {RU(\mathbb{Z}/3)} && {} \\
	\\
	&& {\mho^6_{\text{Spin}}(BL)} & {} \\
	{\mho^6_{\text{Spin}}(B\SL(2,\mathbb{F}_3))} &&&& {\mho^6_{\text{Spin}}(B\Z/9)} \\
	&& {\mho^6_{\text{Spin}}(B\mathbb{Z}/3)}
	\arrow["{p^{\prime\ast}}", from=2-1, to=1-3]
	\arrow["{j^{\prime\ast}}", from=1-3, to=2-5]
	\arrow["{j^\ast}"', from=2-1, to=3-3]
	\arrow["{p^\ast}"', from=3-3, to=2-5]
	\arrow["{p^{\prime\ast}}", from=6-1, to=5-3]
	\arrow["{j^{\prime\ast}}", from=5-3, to=6-5]
	\arrow["{j^\ast}"', from=6-1, to=7-3]
	\arrow["{p^\ast}"', from=7-3, to=6-5]
	\arrow["{\exp(2\pi \ii\eta)}", color={rgb,255:red,153;green,92;blue,214}, curve={height=18pt}, from=2-1, to=6-1]
	\arrow[color={rgb,255:red,153;green,92;blue,214}, curve={height=18pt}, from=3-3, to=7-3]
	\arrow[color={rgb,255:red,153;green,92;blue,214}, curve={height=18pt}, dashed, from=1-3, to=5-3]
	\arrow[color={rgb,255:red,153;green,92;blue,214}, curve={height=18pt}, dashed, from=2-5, to=6-5]
	\arrow["{\pi^{\prime\ast}}"', color={rgb,255:red,214;green,92;blue,92}, curve={height=-18pt}, from=6-5, to=5-3]
	\arrow["{\pi^\ast}", color={rgb,255:red,214;green,92;blue,92}, curve={height=-18pt}, from=7-3, to=6-1]
\end{tikzcd}
\end{equation}

As the irrep $\textbf{1}^{\prime}\in RU(\SL(2,\mathbb{F}_3))$ is pulled back by $j^{*}$ to the irrep $[1]_3\in RU(\Z/3)$, exactly as in \S \ref{sec:A4trunc}, the argument for finding the anomaly $\exp(2\pi \ii \eta)_L(p^{\prime *} (\textbf{1}^{\prime})) \in \mho^6_{\Spin}(BL)$ goes exactly the same way. Thus, we obtain
\begin{equation}
\exp(2\pi \ii \eta)_L(p^{\prime *}(\textbf{1}^{\prime})) =  \exp(2\pi \ii/3)\, .
\end{equation}
We now repeat the argument to find the anomaly of the irrep $\textbf{2}^{\prime}\in RU(\SL(2,\mathbb{F}_3))$ pulled back to the representation $p^{\prime *}(\textbf{2}^{\prime})$ of $L$. Under various maps on the top square of the commutative cube \eqref{eq:SL23-cube}, we get
\begin{equation}
\begin{tikzcd}
	&& {p^{\prime *}(\textbf{2}^{\prime})} \\
	{\textbf{2}^{\prime}} &&&& {[0]_9\oplus [6]_9} \\
	&& {[0]_3 \oplus [2]_3}
	\arrow["{j^\ast}"', from=2-1, to=3-3]
	\arrow["{p^{\prime\ast}}", from=2-1, to=1-3]
	\arrow["{j^{\prime\ast}}", from=1-3, to=2-5]
	\arrow["{\pi^{\prime\ast}}"', color={rgb,255:red,214;green,92;blue,92}, curve={height=-18pt}, from=2-5, to=1-3]
	\arrow["{\pi^\ast}", color={rgb,255:red,214;green,92;blue,92}, curve={height=-18pt}, from=3-3, to=2-1]
	\arrow["{p^\ast}"', from=3-3, to=2-5]
\end{tikzcd}
\end{equation}
where we find $p^{*}([2]_3) = [6]_9$ by requiring that the action of $1 \mod 3 \in \Z/3$ in the representation $[2]_3$, given by $\exp(4\pi\ii/3)$, is the same as the action of $1 \mod 9 \in \Z/9$ in the $p^{*}([1]_3)=[q]_9$ representation given by $\exp(2\pi \ii q/9)$. Under the natural transformation, this commutative square is pushed downward to the bottom square of the cube \eqref{eq:SL23-cube} that reads
\begin{equation}
\begin{tikzcd}
	&& {?} \\
	{-1 \mod 9} &&&& {-1 \mod 3} \\
	&& {-1 \mod 9}
	\arrow["{j^\ast}"', from=2-1, to=3-3]
	\arrow["{p^{\prime\ast}}", from=2-1, to=1-3]
	\arrow["{j^{\prime\ast}}", from=1-3, to=2-5]
	\arrow["{\pi^{\prime\ast}}"', color={rgb,255:red,214;green,92;blue,92}, curve={height=-18pt}, from=2-5, to=1-3]
	\arrow["{\pi^\ast}", color={rgb,255:red,214;green,92;blue,92}, curve={height=-18pt}, from=3-3, to=2-1]
	\arrow["{p^\ast}"', from=3-3, to=2-5]
\end{tikzcd}
\end{equation}
  where the question mark denotes $ \exp(2\pi \ii \eta)_L(p^{\prime\ast}({\bf 2}^\prime))$, the anomaly that we wish to compute. Since $j^{\prime\ast} \circ \pi^{\prime\ast} = \text{id}$ due to the semi-direct product structure of $L$, we can infer from this commutative square that
\begin{equation}
  \exp(2\pi\ii \eta)_L(p^{\prime *}(\textbf{2}^{\prime})) =  \exp(-2\pi \ii/3)\,,
\end{equation}
and the anomaly cancellation condition \eqref{eq:SL2F-4d-anom-cancellation} is truncated to
\begin{equation}
  \label{eq:SL23-anom-cancellation-trunc}
n(\textbf{1}^{\prime})-n(\overline{\textbf{1}^{\prime}}) - n(\textbf{2}^{\prime})+n(\overline{\textbf{2}^{\prime}}) = 0 \mod 3\, .  
\end{equation}

Equivalently, after the group extension, the subgroup of anomaly-free representations becomes $RU_0(\SL(2,\mathbb{F}_3); L, p^\prime)$ generated by
\begin{equation}
\textbf{1}, \quad \textbf{2}, \quad \textbf{3}, \quad \textbf{1}^{\prime}\oplus \overline{\textbf{1}^{\prime}},\quad \textbf{1}^{\prime}\oplus \overline{\textbf{1}^{\prime}},\quad \textbf{1}^{\prime}\oplus \overline{\textbf{2}^{\prime}},\quad \textbf{1}^{\prime\,\oplus 3}.
\end{equation}
Comparing this with the original $RU_0(\SL(2,\mathbb{F}_3))$ whose generators are given in \eqref{eq:RU0SL}, the anomaly truncation quotient is again isomorphic to $\Z/3$, indicating that the order of the anomaly is truncated by a third. 

\subsubsection{Example (continued): anomaly truncation in a flavour physics model} \label{sec:flavour-truncation}

Recall that in \S \ref{sec:anomalous-flavour} we computed the anomaly in an $\SL(2,\F_3)$ symmetry for a fermion representation ${\bf r}_{\mathrm{BSM}}$ utilised in a particular flavour physics model~\cite{Feruglio:2007uu}. The anomaly was found to be the phase $\exp(2\pi \ii/3)$, corresponding to an order 3 anomaly in the group $\mho_6^\Spin(B\SL(2,\F_3))$.

We have just shown that all order 3 anomalies in $\mho_6^\Spin(B\SL(2,\F_3)$ map to zero under the pullback induced by the $\Z/3$ group extension $p^{\prime}: L \twoheadrightarrow \SL(2,\F_3)$. In our notation, we can write
\begin{align}
{\bf r}_{\mathrm{BSM}} &\notin RU_0(\SL(2,\F_3)), \quad \text{but}\\
{\bf r}_{\mathrm{BSM}} &\in RU_0(\SL(2,\mathbb{F}_3); L, p^\prime)\, .
\end{align}
Physically, this means that, while the theory as set out in~\cite{Feruglio:2007uu} is anomalous {\em per se}, it becomes anomaly-free (for the same fermion content) by considering an enlargement of the symmetry to the $\Z/3$ extension $p^\prime: L \twoheadrightarrow \SL(2,\F_3)$, which is otherwise physically indistinguishable at low energies from $\SL(2,\F_3)$.

\section{Summary}
\label{sec:summary}

In this work, we investigated global anomalies of $4$d chiral fermions  charged under some finite non-abelian symmetry groups $G$, using the rigorous tool of the cobordism classification. Motivated in part by their relevance to flavour physics, we considered as examples the alternating group $A_4$, the quaternion group  $Q_8$, the special linear group $\SL(2,\mathbb{F}_3)$, and the symmetric group $S_3$. 

For each of these groups, we determined the associated bordism group $\Omega^{\Spin}_5(BG)$ that classifies all possible anomalies, and then evaluated the $\eta$-invariant to precisely obtain the anomalies for chiral fermions charged in each representation of these groups. Our methods can be adapted to many other finite groups of interest.

When there is a non-trivial anomaly, the order of the anomaly can be further `truncated' to an effective anomaly of lower finite order by passing to an extension of the original symmetry. After putting this notion of `anomaly truncation' on a mathematically sound footing, we worked out the possible anomaly truncations for all the cases that we considered. 
We conclude this paper by summarising our specific results.

The quaternion group $Q_8$ and the dihedral groups $D_{2n}$ (with a special case $S_3= D_6$) are anomaly-free. There are order 9 anomalies for the groups $A_4$ and $\SL(2,\mathbb{F}_3)$ due to chiral fermions in complex representations of these groups; fermions in the real or pseudo-real representations do not contribute to any anomaly. The anomalies can be concisely expressed as anomaly cancellation conditions. Let $n(\textbf{r})$ denote the number of left-handed fermions in the representation $\textbf{r}$ minus the number of right-handed fermions in the same representation. Then the anomaly cancellation condition reads
\begin{equation}
  n(\textbf{1}^{\prime}) - n(\overline{\textbf{1}^{\prime}}) = 0 \mod 9
\end{equation}
for the group $A_4$, and
\begin{equation}
  n(\textbf{1}^{\prime}) - n(\overline{\textbf{1}^{\prime}})-   n(\textbf{2}^{\prime}) + n(\overline{\textbf{2}^{\prime}}) = 0 \mod 9
\end{equation}
for the group $\SL(2,\mathbb{F}_3)$. The labels for the irreps are those listed in the character tables \ref{tab:A4-char} and \ref{tab:SL2F-char}. As an example, we show that the flavour physics model in Ref.~\cite{Feruglio:2007uu} has an order 3 anomaly.

Both $A_4$ and $\SL(2,\mathbb{F}_3)$ can be extended by $\Z/3$ to $(\Z/2\times \Z/2) \rtimes \Z/9$ and $Q_8 \rtimes \Z/9$, respectively, with the effect of truncating these order $9$ anomalies  down to order $3$ anomalies. The anomaly cancellation conditions above remain the same apart from being taken modulo $3$ instead of modulo $9$. Under such a symmetry extension, the flavour physics model of Ref.~\cite{Feruglio:2007uu} thus becomes anomaly-free. 

\acknowledgments 

We are grateful to Arun Debray for divulging some helpful tricks for computing spectral sequences.
JD has received funding from the European Research Council (ERC) under the European Union’s Horizon 2020 research and innovation programme under grant agreement 833280 (FLAY), and by the Swiss National Science Foundation (SNF) under contract 200020-204428. BG is supported by STFC consolidated grant ST/T000694/1. NL is supported by an STFC consolidated grant in ‘Particles, Strings and Cosmology', and by the Royal Society.

\appendix

\section{Spin bordism groups $\Omega^{\Spin}_d(BG)$ for various non-abelian
  finite group $G$}
\label{app:bordism-results}

In this Appendix, we collect facts about the non-abelian discrete groups considered in the main part of the paper.

There is an isomorphism between the spin bordism group and the connective real K-theory,
\begin{equation}
\Omega^{\Spin}_d(X) \cong ko_d(X),
\end{equation}
when $d\le 7$, by virtue of the Anderson--Brown--Peterson theorem \cite{ABP:1967} which proves that the Atiyah--Bott--Shapiro map \cite{ATIYAH19643}  $MT\Spin \to ko$ is 7-connected  \cite{Debray:2021rik}. Here, $ko$ is the connective real K-theory spectrum and $MT\Spin$ is the stable version of the Madsen--Tillmann spectrum of the spin group which represent the $ko$-homology and the spin bordism, respectively, according to Brown's representability theorem \cite{Brown:1962, ADAMS1971185}. There are already many computations and results on the connective real K-theory of a space $X$ when $X$ is the classifying space of a discrete group, which are nicely summarised in Ref. \cite{bruner2010connective}. The bordism groups quoted in this Appendix will be mainly from this reference.

When there is an ambiguity in the existing $ko$-computations, or when there is no result at all, we make use of both the Atiyah--Hirzebruch spectral sequence (AHSS) and the Adams spectral sequence (ASS).
For an introduction to the use of these spectral sequences for calculating bordism groups in physics context, see {\em e.g.} Refs. \cite{Garcia-Etxebarria:2018ajm,Davighi:2019rcd} for the AHSS, and Ref. \cite{beaudry2018guide} for the ASS.

\subsection{$G=\Z/2 \times \Z/2$}
\label{app:Z2Z2}

The $ko$-homology groups for $BG$ when $G=\Z/2 \times \Z/2$ were first
computed in Ref. \cite{YuPHD:1995} (which computed the $ko$-homology
groups more generally for $BG$ when $G=(\Z/2)^k$). The calculations
were also used in physics literature in recent years (see
{\em e.g.} \cite{Guo:2018vij}).

We first decompose $B\Z/2 \times B\Z/2$ into a wedge sum of different
factors as
\begin{equation}
B\Z/2 \times B\Z/2 = (B\Z/2) \vee (B\Z/2) \vee \left( B\Z/2 \wedge B\Z/2 \right) \nn
\end{equation}
and get
\begin{equation}
\widetilde{ko}_d(B\Z/2 \times B\Z/2) \cong \widetilde{ko}_d(B\Z/2) \oplus \widetilde{ko}_d(B\Z/2) \oplus \widetilde{ko}_d((B\Z/2)^{\wedge 2}),
\end{equation}
where we use the notation
\begin{equation}
X^{\wedge k} = \underbrace{X \wedge \ldots \wedge X}_{k\;\text{times}}
\end{equation}
to denote the $k$-fold wedge sum of $X$.
$\widetilde{ko}_d((B\Z/2)^{\wedge k})$ were calculated in
Ref.~\cite{YuPHD:1995} to be
\begin{equation}
\label{eq:koBZ22}
\widetilde{ko}_d((B\Z/2)^{\wedge k}) \cong
\begin{cases}
  \Z/2 \times (\Z/2)^{C(k,d)}, & \qquad d=8l+1, \quad k<4l+2,\\
  & \qquad d=8l+2, \quad k < 4l+3,\\
  \Z/2^{4l+4-k} \times (\Z/2)^{C(k,d)}, & \qquad d=8l+3, \quad k < 4l+4,\\
  \Z/2^{4l+5-k} \times (\Z/2)^{C(k,d)}, & \qquad d=8l+7, \quad k < 4l+5,\\
  (\Z/2)^{C(k,d)}, & \qquad \text{otherwise}.
\end{cases}
\end{equation}
Here, $C(k,d)$ denotes the coefficient of the monomial $t^d$ in the
infinite series
\begin{equation}
P_k(t) = \frac{t^k \left( 1-(1-t)^{k-1} \right) Q_k(t)}{\left( 1-t^4 \right) \left( 1+t^3 \right) \left( 1-t \right)^{k-1}},\nn
\end{equation}
where
\begin{equation}
Q_k(t) =
\begin{cases}
  t^{k-1}, & \qquad k=0,1 \mod 4,\\
  t^{k-2}\left( 1+t-t^2+t^3-t^5 \right), & \qquad k = 2 \mod 4,\\
  t^{k-3}\left( 1+t^{3-t^5} \right), & \qquad k=3 \mod 4.
\end{cases}\nn
\end{equation}

Combining all the results, specialising to $k=2$ and using the
isomorphism between $\Omega^{\Spin}$ and $ko$, the reduced spin bordism
groups of $B\Z/2 \times B\Z/2$  are given in Table
\ref{tab:Om-BZ2BZ2} (see also Fig. 12.3.3 of
Ref. \cite{bruner2010connective}). \begin{table}[h]
  \centering
  \begin{tabular}{|c|ccccccc|}
    \hline
    $d$ & $0$ & $1$ & $2$ & $3$ & $4$ & $5$ & $6$\\
    \hline
    $\tilde{\Omega}^{\Spin}_d(B\Z/2 \times B\Z/2)$ & $0$ & $(\Z/2)^2$ & $(\Z/2)^3$ & $\Z/4 \times (\Z/8)^2$ & $(\Z/2)^2$ & $0$ & $\Z/2$\\
    \hline
  \end{tabular}
  \caption{The reduced spin bordism groups of $B\Z/2 \times B\Z/2$ up to degree $6$}
  \label{tab:Om-BZ2BZ2}
\end{table}

\subsection{$G=A_4$} \label{app:A4}

Consider $A_4$, the alternating group of order $4$. Its derived
subgroup $A_4^{\prime}$ is $\Z/2 \times \Z/2$, giving rise to the short exact
sequence
\begin{equation}
  \label{eq:A4-SES}
0 \to \Z/2 \times \Z/2 \to A_4 \to \Z/3 \to 0.
\end{equation}
and take $X$ to be its classifying space $BG$. As outlined above, we
obtain the spin bordism groups $X$ through the isomorphism with the
$ko$-homology groups of $X$, whose calculations can be found in more
quantities in mathematics literature. $ko_d(BA_4)$ have been computed
in Ref.~\cite{bruner2010connective}, with results nicely summarised in
§7.7.E. therein. Using the isomorphism to write their results as spin
bordism and leaving out the factors corresponding to the spin bordism
groups of a point, $\rbord_d(\text{pt})$, the reduced spin bordism
groups of $BA_4$ are given by
\begin{equation}
  \label{eq:BA4-bord}
  \begin{split}
    \rbord_1(BA_4) &\cong \Z/3,\\
    \rbord_2(BA_4) &\cong \Z/2,\\
    \rbord_3(BA_4) &\cong \Z/12,\\
    \rbord_4(BA_4) &= 0,\\
    \rbord_5(BA_4) &\cong \Z/9,\\
	\rbord_6(BA_4) &\cong \Z/2.
  \end{split}
\end{equation}

\subsection{The quaternion group $G= Q_8$}
\label{app:Q8}

Let's now consider the quaternion group $Q_8$, a multiplicative group
of order $8$ whose elements are the basic quaternions $1,i,j,k$
together with their additive inverses $-1,-i,-j,-k$. Its derived
subgroup $Q_8^{\prime}$ is isomorphic to $\Z/2$, and
$Q_8/Q_8^{\prime}$ is isomorphic to the Klein four-group, putting
$Q_8$ inside the short exact sequence
\begin{equation}
0 \to \Z/2 \to Q_8 \to \Z/2 \times \Z/2 \to 0.
\end{equation}

The calculations for $\widetilde{ko}_\bullet(BQ_8)$ were carried out in
Ref. \cite{Bayen:1996}. The authors of \cite{Bayen:1996} first
decomposed $BQ_8$ stably as \cite{MARTINO199113}
\begin{equation}
BQ_8 \cong B\SL(2,\mathbb{F}_3) \vee \Sigma^{-1} \left( BS^3/BN \right) \vee \Sigma^{-1} \left( BS^3/BN \right),
\end{equation}
again localised at the prime $2$. Here $B\SL(2,\mathbb{F}_3)$ is the
classifying space of $\SL(2,\mathbb{F}_3)$, the group of $2\times2$ unit
determinant matrices with entries in the finite field $\mathbb{F}_3$,
while $N$ is the normaliser of the maximal torus in $S^3$ (whose group
structure is inherited from the quaternion multiplication, and can be
thought of as the group $\SU(2)$). They then calculated the $ko$
theory (localised at $2$) for each factor, as given in
Eqs. \eqref{eq:koBSL2F3} and \eqref{eq:koBS3BN} with $n\geq 0$:
\begin{equation}
\label{eq:koBSL2F3}
\widetilde{ko}_{\epsilon+8n}B\SL_2(\mathbb{F}_3) \cong
\begin{cases}
  \Z/2^{4n+3} \oplus \Z/2^{2n},& \qquad \epsilon=3,\\
  \Z/2,& \qquad \epsilon=4,6,8,10,\\
  \Z/2 \oplus \Z/2,& \qquad \epsilon=5,9,\\
  \Z/2,& \qquad \epsilon=6,\\
  \Z/2^{4n+6} \oplus \Z/2^{2n},& \qquad \epsilon=7.
\end{cases}
\end{equation}
\begin{equation}
\label{eq:koBS3BN}
\widetilde{ko}_{\epsilon+8n} \left( \Sigma^{-1}BS^3/BN \right) \cong
\begin{cases}
  \Z/2, & \qquad \epsilon=1,2,\\
  \Z/2^{2n+2}, & \qquad \epsilon=3,7,\\
  0, & \qquad \epsilon=4,5,6,8.
\end{cases}
\end{equation}
We combine these results and summarise the spin bordism groups of
$BQ_8$ in Table \ref{tab:ko-BQ8} for convenience (see also §7.4.C of
Ref. \cite{bruner2010connective}). From this identification, we can
deduce that
\begin{equation}
\Omega^{\Spin}_5(BQ_8) \cong \Z/2 \times \Z/2.
\end{equation}
\begin{table}[h]
  \centering
  \begin{tabular}{|c|cccccc|}
    \hline
    $n$ & $1$ & $2$ & $3$ & $4$ & $5$ & $6$ \\
    \hline
    $\widetilde{\Omega}^{\Spin}_n(BQ_8)$ & $(\Z/2)^2$ & $(\Z/2)^2$ & $\Z/8 \times (\Z/4)^2$ & $\Z/2$ & $(\Z/2)^2$ & $\Z/2$\\
    \hline
  \end{tabular}
  \caption{The reduced spin bordism groups for $BQ_8$ in degrees lower than $7$.}
  \label{tab:ko-BQ8}
\end{table}

\subsection{$G=\SL(2,\mathbb{F}_3)$}
\label{app:SL2F3}

The group
\begin{equation}
\SL(2,\mathbb{F}_3) := \left\{
  \begin{pmatrix}
    a & b\\ c & d 
  \end{pmatrix} \bigg|\, a,b,c,d \in \mathbb{F}_3 , \,ad-bc=1
\right\}
\end{equation}
is a group extension of $\Z/3$ by the quaternion group $Q_8$,
\begin{equation} \label{eq:SL2Fextension}
0 \to Q_8 \to \SL(2,\mathbb{F}_3) \to \Z/3 \to 0.
\end{equation}

\begin{figure}[h]
  \centering
  \includegraphics[scale=0.75]{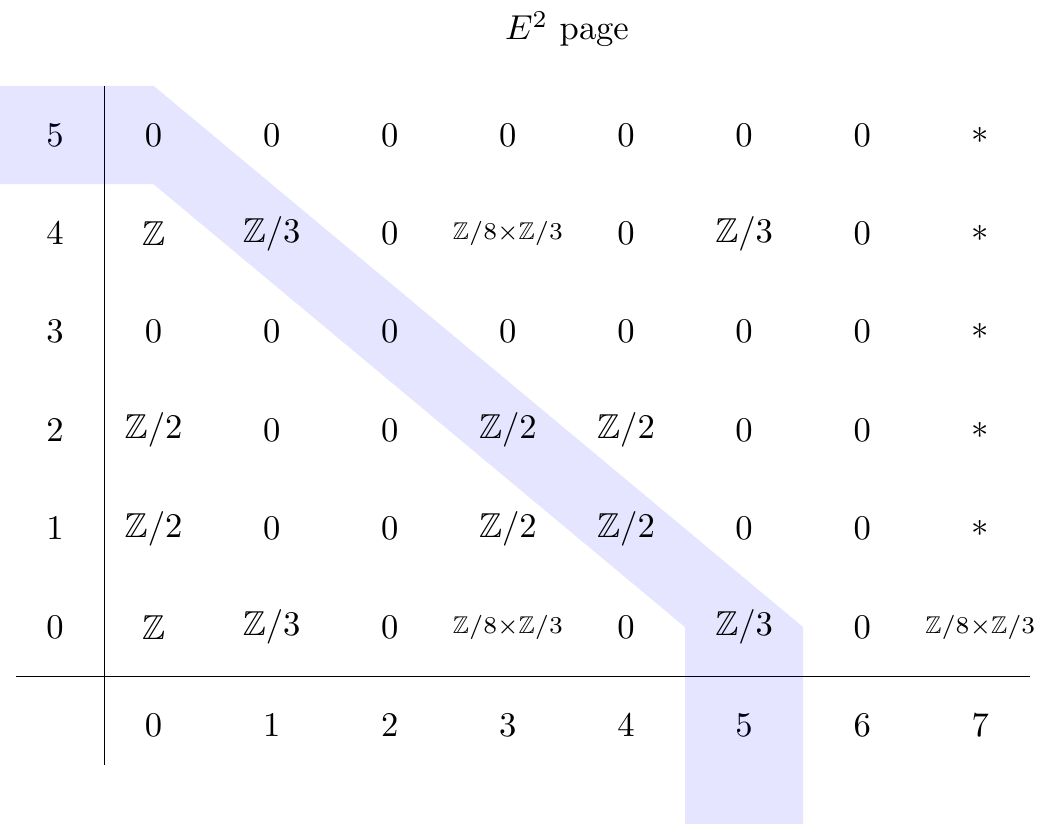}
  \caption{The $E^2$ page for the Atiyah--Hirzebruch spectral sequence calculating $\Omega^\Spin_\bullet(B\SL(2,\mathbb{F}_3))$ .}
  \label{fig:AHSS-SL23}
\end{figure}

We know from \cite{Bayen:1996} that at 2-localisation,
\begin{equation}
\widetilde{ko}_5(B\SL(2,\mathbb{F}_3))^{\wedge}_2  \cong \Z/2 \times \Z/2,
\end{equation}
but since there is a $3$-torsion present in $\SL(2,\mathbb{F}_3)$,
there is a possibility that there is a non-trivial odd-torsion piece
in the spin bordism of $B\SL(2,\mathbb{F}_3)$. To find the odd
torsion, we consider the Atiyah--Hirzebruch spectral sequence for the
fibration $\text{pt} \to B\SL(2,\mathbb{F}_3) \to B\SL(2,\mathbb{F}_3)$,
\begin{equation}
  \label{eq:AHSS-SL23}
E^2_{p,q} = H_p(B\SL(2,\mathbb{F}_{3}),\Omega^{\Spin}_q(\text{pt})) \Rightarrow \Omega^{\Spin}_{p+q}(B\SL(2,\mathbb{F}_3)) .
\end{equation}
To construct the entries on the $E^2$ page, we need to know both the
integral and mod 2 homology groups of $B\SL(2,\mathbb{F}_3)$. Using
the isomorphism between the singular homology of $BG$ and the group homology of $G$, $H_{\bullet}(BG;\Z) \cong \mathcal{H}_{\bullet}(G,\Z)$, with help from \texttt{GAP} \cite{GAP4}, we obtain the homology groups of $B\SL(2,\mathbb{F}_3)$ as shown in Table
\begin{table}[h]
  \centering
  \begin{tabular}{|c|cccccccc|}
    \hline
    $n$ & $0$ & $1$ & $2$ & $3$ & $4$ & $5$ & $6$ & $7$ \\
    \hline
    $H_n(B\SL(2,\mathbb{F}_3);\Z)$ & $\Z$ & $\Z/3$ & $0$ & $\Z/8 \times \Z/3$ & $0$ & $\Z/3$ & $0$ & $\Z/8 \times \Z/3$\\
    $H_n(B\SL(2,\mathbb{F}_3);\Z/2)$ & $\Z/2$ & $0$ & $0$ & $\Z/2$ & $\Z/2$ & $0$ & $0$ & $\Z/2$\\
    \hline
  \end{tabular}
  \caption{The homology groups of $B\SL(2,\mathbb{F}_3)$ with coefficients in $\Z$ and $\Z/2$ up to degree $7$, as we need to compute the bordism groups up to degree $6$.}
  \label{tab:hom-SL23}
\end{table}
The $E^2$ page of the AHSS \eqref{eq:AHSS-SL23} can be constructed as shown in Fig. \ref{fig:AHSS-SL23}.
The entries on the diagonal $p+q=5$ already stabilise on this page. To determine the 5th bordism group, however, we need to solve the non-trivial extension problem. Using the result from the 2-localisation above, we narrow down the extensions to two options:
\begin{equation} \label{eq:SL2Foptions}
\Omega^{\Spin}_5(B\SL(2,\mathbb{F}_3)) \cong \Z/2 \times \Z/2 \times \Z/9 \quad \text{or} \quad \Z/2 \times \Z/2 \times \Z/3 \times \Z/3 .
\end{equation}

To eliminate this ambiguity, we apply a homological version of the `anomaly interplay' procedure used in {\em e.g.} \S~\ref{sec:A4-4d}. Firstly, the extension sequence (\ref{eq:SL2Fextension}) is right-split, meaning that $\SL(2,\mathbb{F}_3)$ is a semi-direct product
\begin{equation}
\SL(2,\mathbb{F}_3) = Q_8 \rtimes \Z/3,
\end{equation}
so in addition to the quotient map $\pi:\SL(2,\mathbb{F}_3) \to \Z/3$, we also have the subgroup embedding map $i:\Z/3 \to \SL(2,\mathbb{F}_3)$ such that $\pi \circ i$ is the identity. Applying the bordism (covariant) functor $\Omega^\Spin_\bullet(B\cdot)$, we obtain maps
\begin{equation}
\begin{tikzcd}
	{\mathbb{Z}/9} & {\Omega_5^\text{Spin}(B\text{SL}(2,\mathbb{F}_3))} & {\mathbb{Z}/9}
	\arrow["{\pi_*}", from=1-2, to=1-3]
	\arrow["{i_*}", from=1-1, to=1-2]
	\arrow["{\text{id}}"', curve={height=18pt}, dashed, from=1-1, to=1-3]
\end{tikzcd}\, ,
\end{equation}
with the composition $\pi_{*} \circ i_{*}$ being the identity. Thus, the pushforward map $i_{*}$ must factor through a $\Z/9$ subgroup inside $\Omega_5^\text{Spin}(B\SL(2,\mathbb{F}_3))$, and from the two options (\ref{eq:SL2Foptions}), we deduce that
\begin{equation}
\Omega^{\Spin}_5(B\SL(2,\mathbb{F}_3)) \cong \Z/2 \times \Z/2 \times \Z/9.
\end{equation}
The reduced spin bordism groups in degrees $\leq 5$ are easily read off from the $E^2$ page of Fig.~\ref{fig:AHSS-SL23}, and recorded in Table~\ref{tab:bord-results}.
From Fig.~\ref{fig:AHSS-SL23}, we see that the 6th bordism group depends on an unknown differential $d^3:E^3_{7,0} \to E^3_{4,2}:\Z/8 \times \Z/3 \mapsto \Z/2$. Without further information, this map could be either zero or non-zero, in which case the $E_{4,2}$ element stabilises either to $\Z/2$ or $0$ respectively. Thus we learn from the AHSS that $\Omega^{\Spin}_6(B\SL(2,\mathbb{F}_3)) \cong \Z/2$ or $0$. Computation of $\text{ko}(B\SL(2,\mathbb{F}_3))$ at 2-completion in Ref. \cite{Bayen:1996}, which we have quoted in Eq. \eqref{eq:koBSL2F3}, tells us that this differential must be trivial and $\Omega^{\Spin}_6(B\SL(2,\mathbb{F}_3))$ must be isomorphic to $\Z/2$.

\subsection{The dihedral group $G=D_{2n}$}
\label{app:D2n}

In this Appendix, we will show that $\Omega^{\Spin}_5(BD_{2n})$ vanishes for any $n$.

\subsubsection*{Odd $n$}

\begin{figure}[h]
  \centering
  \begin{subfigure}[b]{0.45\textwidth}
    \centering
  \includegraphics[scale=0.5]{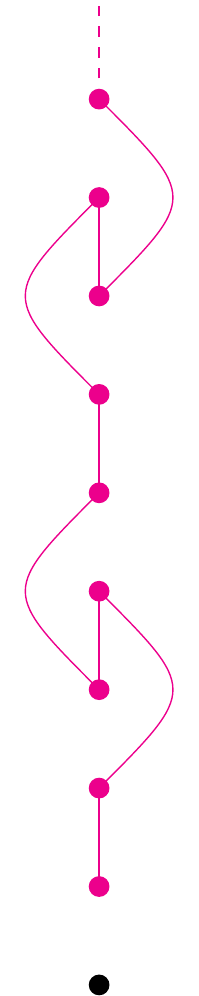}
  \caption{The structure of $H^\bullet(BD_{2n};\Z/2)$ as an $\mathcal{A}(1)$-module for odd $n$}
  \label{fig:BD2odd-cells}
\end{subfigure}
\hfill
\begin{subfigure}[b]{0.45\textwidth}
  \centering
  \includegraphics[scale=0.6]{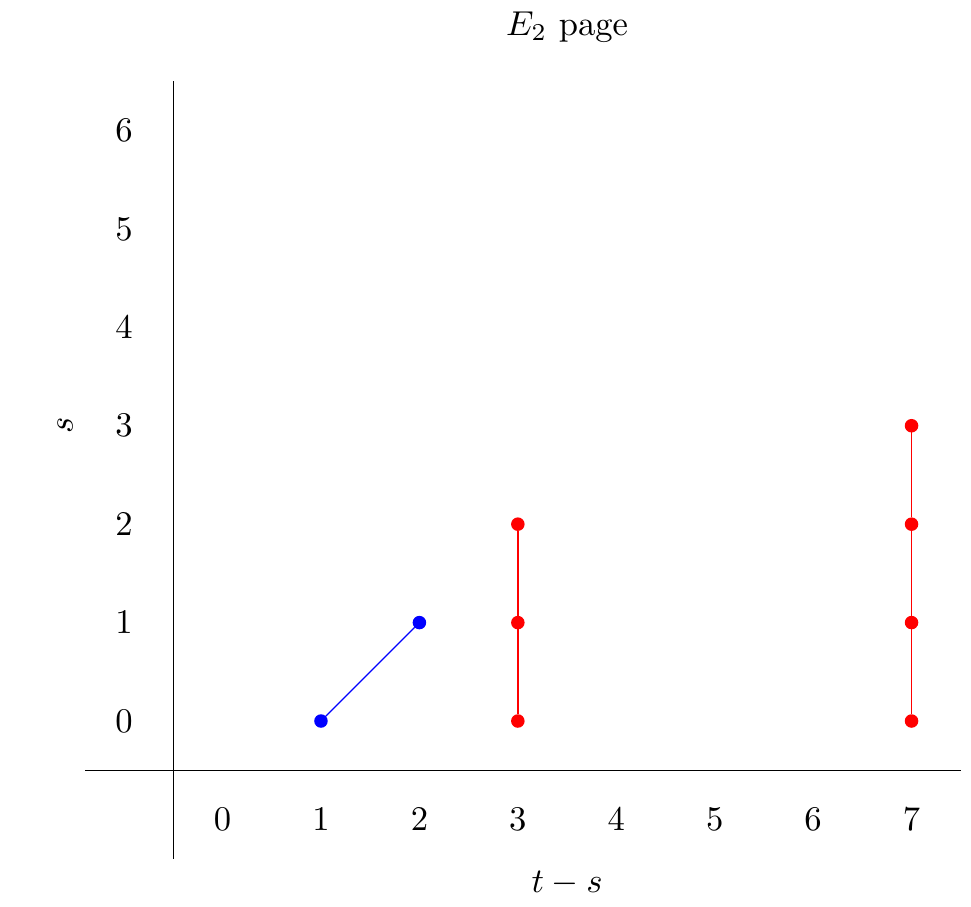}
    \caption{The Adams chart for $\ext^{s,t}_{\mathcal{A}(1)}(\tilde{H}^{\bullet}(BD_{2n};\Z/2),\Z/2)$}
    \label{fig:ASS-BD2odd}
  \end{subfigure}
  \caption{}
\end{figure}

Consider first the case with odd $n$. The integral cohomology ring and the mod 2 cohomology ring of $BD_{2n}$ are given by \cite{Handel:1993}
\begin{equation}
\label{eq:D2odd-cohom-rings}
H^{\bullet}\left( BD_{2n};\Z \right) \cong \frac{\Z[a_2,d_4]}{(2a_2,nd_4)}, \qquad H^{\bullet} \left( BD_{2n};\Z/2 \right) \cong \Z/2[u_1],
\end{equation}
where the subscript on each indeterminate denotes its degree. We will now apply the Adams spectral sequence
\begin{equation}
  \label{eq:Adams-SS-BD2n}
E^{s,t}_2 = \ext^{s,t}_{\mathcal{A}(1)}(\tilde{H}^{\bullet}(BD_{2n};\Z/2),\Z/2) \Rightarrow \tilde{\Omega}^{\Spin}_{t-s}(BD_{2n})^{\wedge}_2, \quad t-s\le 7,
\end{equation}
to find 2-torsion factors of the spin bordism groups of $BD_{2n}$. As an $\mathcal{A}(1)$-module, $H^{\bullet}(BD_{2n};\Z/2)$ has the structure as depicted by a cell diagram in Fig.~\ref{fig:BD2odd-cells}.
Discarding the summand $\Z/2$ which is not part the reduced mod 2 cohomology, we obtain the Adams chart for $\ext^{s,t}_{\mathcal{A}(1)}(\tilde{H}^{\bullet}(BD_{2n};\Z/2),\Z/2)$ as shown in Fig.~\ref{fig:ASS-BD2odd},
whence the 2-torsion factors in the spin bordism groups of $BD_{2n}$ can be read off:
\begin{equation}
  \begin{split}
    &\tilde{\Omega}^{\Spin}_1(BD_{2n})^{\wedge}_2 = \tilde{\Omega}^{\Spin}_2(BD_{2n})^{\wedge}_2 \cong \Z/2,\quad \tilde{\Omega}^{\Spin}_3(BD_{2n})^{\wedge}_2 \cong \Z/8, \\
    &\tilde{\Omega}^{\Spin}_4(BD_{2n})^{\wedge}_2 = \tilde{\Omega}^{\Spin}_5(BD_{2n})^{\wedge}_2  = \tilde{\Omega}^{\Spin}_6(BD_{2n})^{\wedge}_2  = 0.
  \end{split}
\end{equation}

\begin{figure}[h]
  \centering
  \includegraphics[scale=0.7]{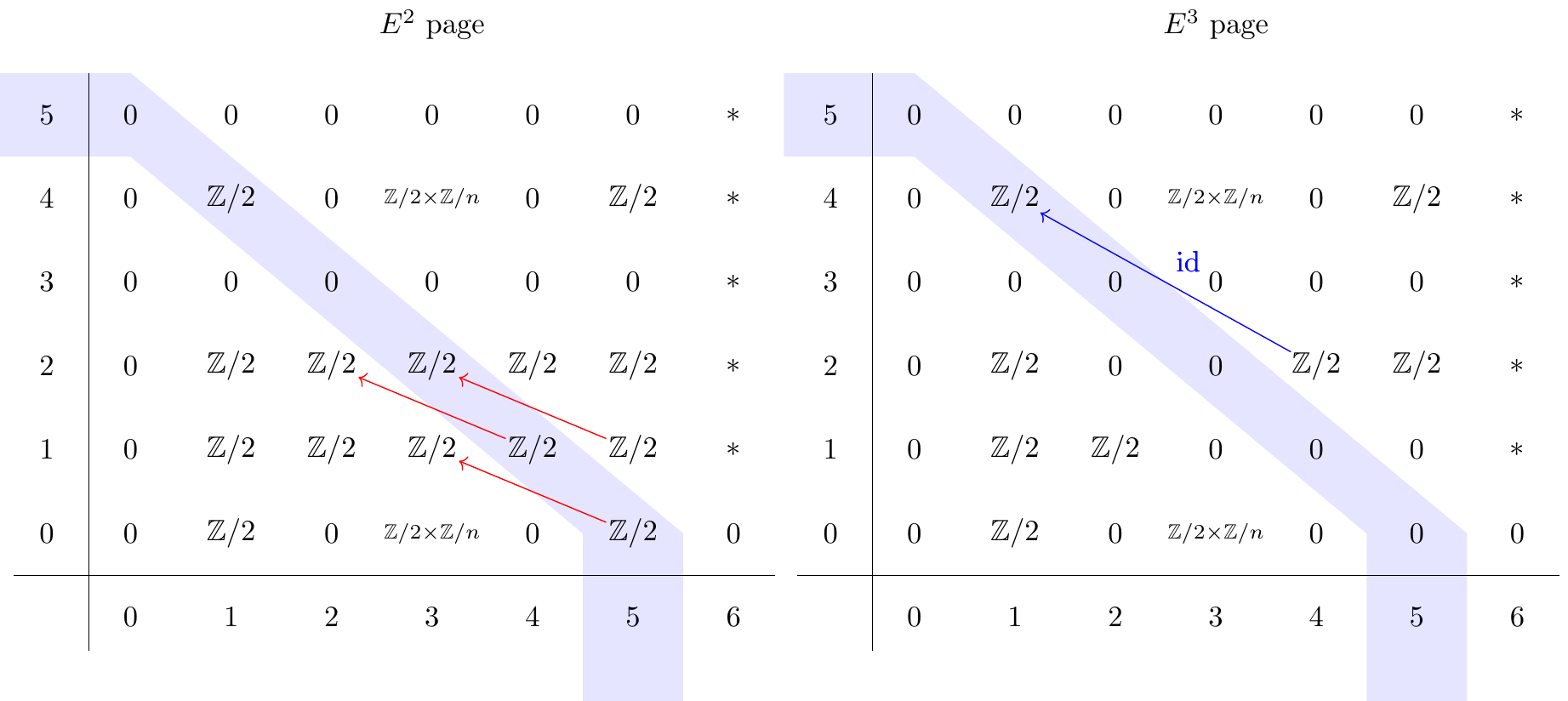}
  \caption{Pages $E^2$ and $E^3$ of the AHSS for $\tilde{\Omega}^\Spin_d(BD_{2n})$ with odd $n$}
  \label{fig:AHSS-BD2odd}
\end{figure}

We complete our computation by working out the possible odd-torsion through the Atiyah--Hirzebruch spectral sequence (AHSS). Applying the universal coefficient theorem to the cohomology groups given in Eq.~\eqref{eq:D2odd-cohom-rings}, we obtain the homology groups, as shown in Table \ref{tab:hom-BD2odd}.
\begin{table}[h]
  \centering
  \begin{tabular}{|c|cccccccc|}
    \hline
    $i$ & $0$ & $1$ & $2$ & $3$ & $4$ & $5$ & $6$ & $7$\\
    \hline
    $H_i(BD_{2n};\Z)$ & $\Z$ & $\Z/2$ & $0$ & $\Z/2 \times \Z/n$ & $0$ & $\Z/2$ & $0$ & $\Z/n$\\
    $H_i(BD_{2n};\Z/2)$ & $\Z/2$ & $\Z/2$ & $\Z/2$ & $\Z/2$ & $\Z/2$ & $\Z/2$ & $\Z/2$ & $\Z/2$\\
    \hline
  \end{tabular}
  \caption{Integral and mod 2 homology groups of $BD_{2n}$ with odd $n$.}
  \label{tab:hom-BD2odd}
\end{table}
We can then use this data to construct the corresponding AHSS for the
fibration $\text{pt} \to BD_{2n} \to BD_{2n}$ converging to the reduced bordism
groups,
\begin{equation}
E^2_{p,q} = \tilde{H}_p(BD_{2n}; \Omega^{\Spin}_q(\text{pt})) \Rightarrow \widetilde{\Omega}^{\Spin}_{p+q}(BD_{2n}),
\end{equation}
as shown in Fig. \ref{fig:AHSS-BD2odd}. The non-trivial differentials on
the $E^2$ page are denoted by the red arrows. These differentials can
be deduced from the fact that the differentials $d_2$ acting on the
$0$th and $1$st rows in the $E^2$ page of the AHSS for the fibration
$\text{pt} \to BG \to BG$ converging to a spin bordism group is dual to
the action of the Steenrod square $\sq^2$ on the mod 2 cohomology
\cite{teichner1992topological,teichner1993signature}. The mod 2
cohomology ring of $BD_{2n}$ with $n$ odd is generated by a single generator of degree
one, so the action of the Steenrod squares on any element of the
cohomology ring can be deduced from the axioms.

The entries with $p+q\le 6$ all stabilise by the $E^4$ page, where we have used the consistency with the Adams spectral sequence calculation above to deduce that $d_3:E^3_{4,2}\to E^3_{1,4}$ is the identity map. The computation shows us that for $d\leq 6$, there is an odd-torsion factor in $\Omega^{\Spin}_d(BD_{2n})$ only when $d=3$:
\begin{equation}
\Omega^{\Spin}_3(BD_{2n}) \cong \Z/n \times \Z/8, \qquad n \quad\text{odd}.
\end{equation}
 The resulting spin bordism groups of $BD_{2n}$, with odd $n$, are summarised in Table \ref{tab:bord-results}.

 \subsubsection*{Even $n$}
 
\begin{figure}[h]
  \centering
  \begin{subfigure}[b]{0.45\textwidth}
    \centering
  \includegraphics[scale=0.45]{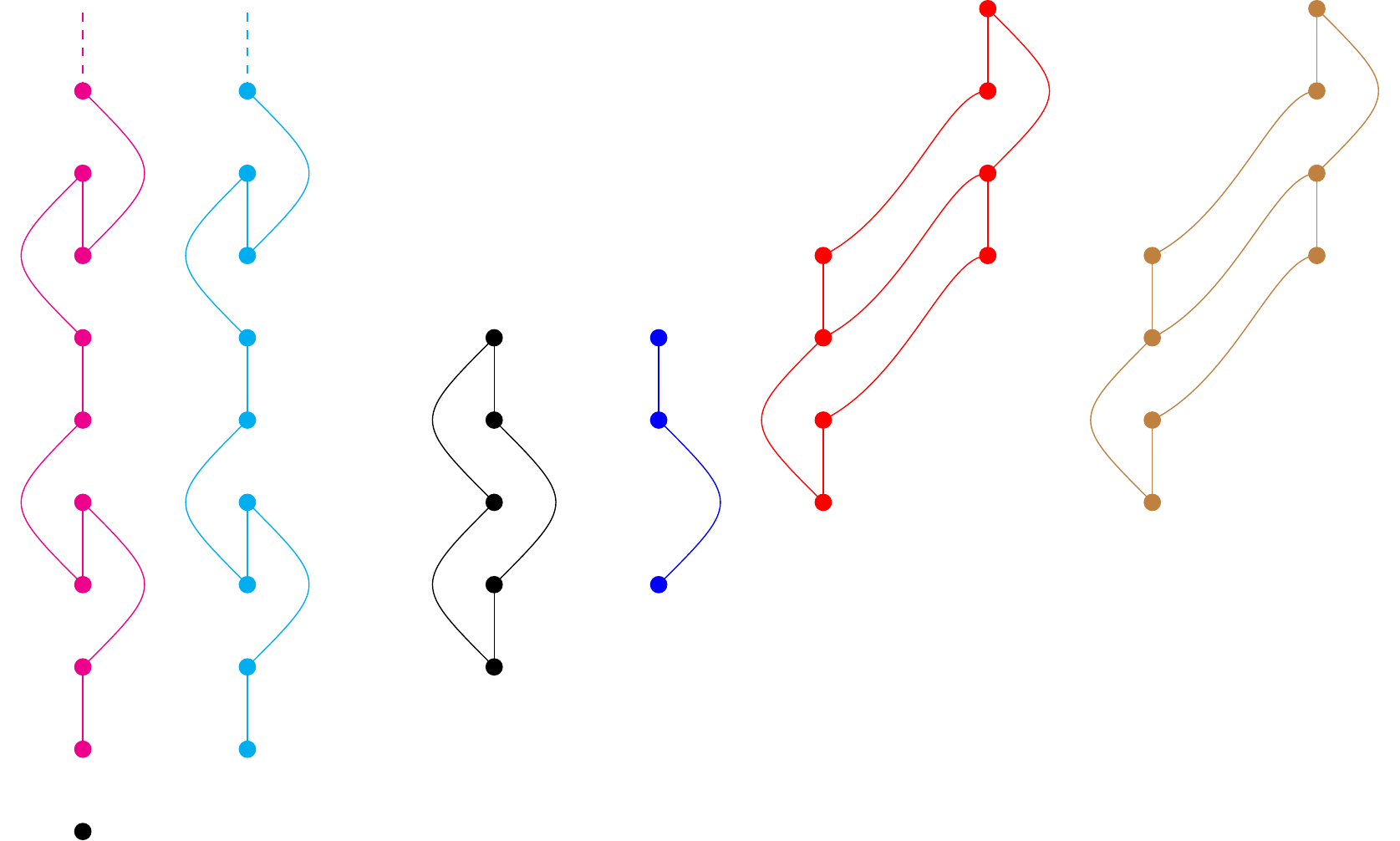}
  \caption{The structure of $H^\bullet(BD_{2^{k+2}};\Z/2)$ as an $\mathcal{A}(1)$-module}
  \label{fig:BDeven-A1mod}
\end{subfigure}
\hfill
\begin{subfigure}[b]{0.45\textwidth}
  \centering
  \includegraphics[scale=0.5]{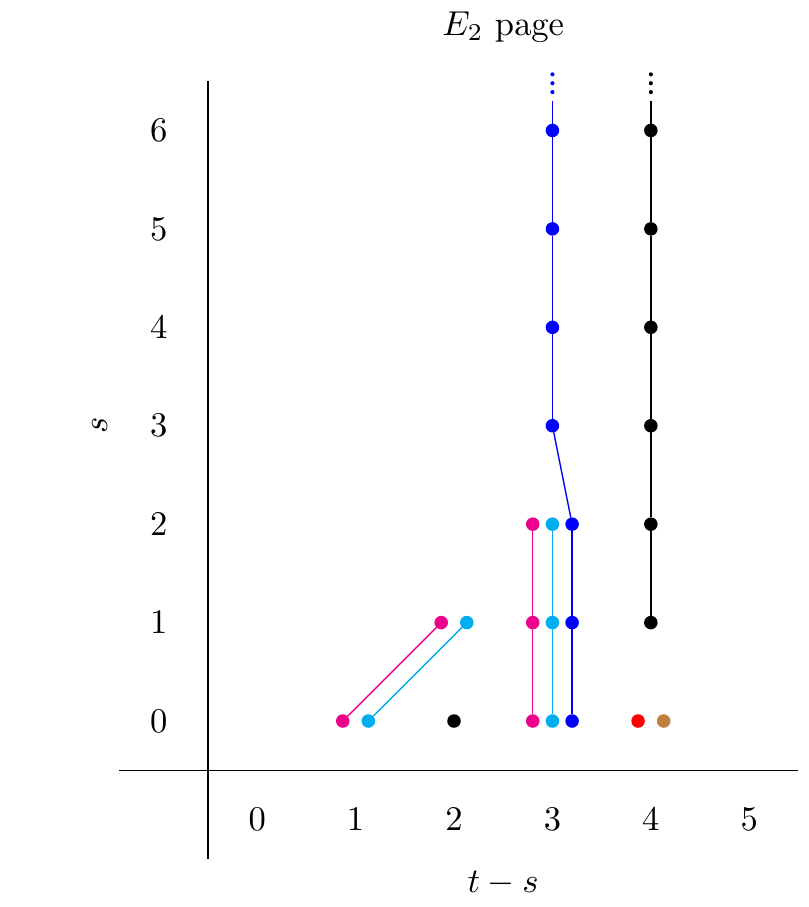}
  \caption{The Adams chart for $\ext^{s,t}_{\mathcal{A}(1)}\left(\tilde{H}^{\bullet}(BD_{2^{k+2}};\Z/2),\Z/2\right)$}
  \label{fig:AdamsSS-BD2n}
\end{subfigure}
\caption{}
\end{figure}

The calculation of the spin bordism groups of $BD_{2n}$ when $n$ is even is more involved. For $n$ even, the cohomology ring is given by
\cite{Handel:1993}
\begin{equation}
H^{\bullet}(BD_{2n};\Z) \cong \Z[a_2,b_2,c_3,d_4]/ I,
\end{equation}
where the subscripts on the generators denote the degree, and $I$ is
the ideal generated by $2a_2$, $2b_2$, $2c_3$, $nd_4$,
$b_2^2+a_2b_2+(n^2/4)d_4$, and $c_3^2+a_2d_4$. Writing this out
explicitly, we have (for $p\ne 0$)
\begin{equation}
H^p(BD_{2n};\Z) \cong
\begin{cases}
  (\Z/2)^{\frac{p-1}{2}}, & \qquad p=1,3 \mod 4\\
  (\Z/2)^{\frac{p+2}{2}}, & \qquad p =2 \mod 4\\
  \Z/n \times (\Z/2)^{p/2}, & \qquad p = 0 \mod 4
\end{cases}
\end{equation}
The mod 2 cohomology ring is given by \cite{Handel:1993}
\begin{equation}
  \label{eq:mod2-cohom-D2even}
  H^{\bullet} \left( BD_{2n};\Z/2 \right) \cong \frac{\Z/2[u_1,v_1,w_2]}{\left( u_1^2+u_1v_1+ (n/2)w_2 \right)}, \quad n\,\text{even},
\end{equation}
with $\sq^1(w_2)=v_1w_2$ (the action of the Steenrod squares on other generators follows automatically from the axioms). Using the same strategy as the odd $n$ case, we first construct the Adams spectral sequence \eqref{eq:Adams-SS-BD2n} to find the 2-torsion factors. As a $\mathcal{A}(1)$-module, the structure of $H^{\bullet}(BD_{2n};\Z/2)$ is given in Fig. \ref{fig:BDeven-A1mod}. Ignoring the summand $\Z/2$ which gives us the spin bordism groups of a point, the corresponding the Adams chart for $\ext^{s,t}_{\mathcal{A}(1)}\left(\tilde{H}^{\bullet}(BD_{2^{k+2}};\Z/2),\Z/2\right)$ is shown in Fig. \ref{fig:AdamsSS-BD2n}.
We can see right away that
\begin{equation}
\tilde{\Omega}^{\Spin}(BD_{2n})^{\wedge}_2 \cong \Z/2\times \Z/2, \quad\tilde{\Omega}^{\Spin}_5(BD_{2n})^{\wedge}_2 = 0.
\end{equation}
However, we cannot determine the differentials from the column $t-s=4$ to the $t-s=3$ in general, nor can we solve the extension problem in the column $t-s=2$. Finally, the AHSS for the fibration $\text{pt} \to BD_{2n} \to BD_{2n}$,
\begin{equation}
E^2_{p,q} = H_p(BD_{2n};\Omega^{\Spin}_q(\text{pt})) \Rightarrow \Omega^{\Spin}_{p+q}(BD_{2n}),
\end{equation}
whose $E^2$ page is shown in Fig. , tells us that there is no odd-torsion in the spin bordism group of $BD_{2n}$ of degree $d=5$. Hence, we can conclude that
\begin{equation}
\tilde{\Omega}^{\Spin}_5(BD_{2n}) = 0,
\end{equation}
for any integer $n$. Similarly, we also know that $\tilde{\Omega}^{\Spin}_1(BD_{2n})=(\Z/2)^2$.
\begin{figure}[h]
  \centering
  \includegraphics[scale=0.6]{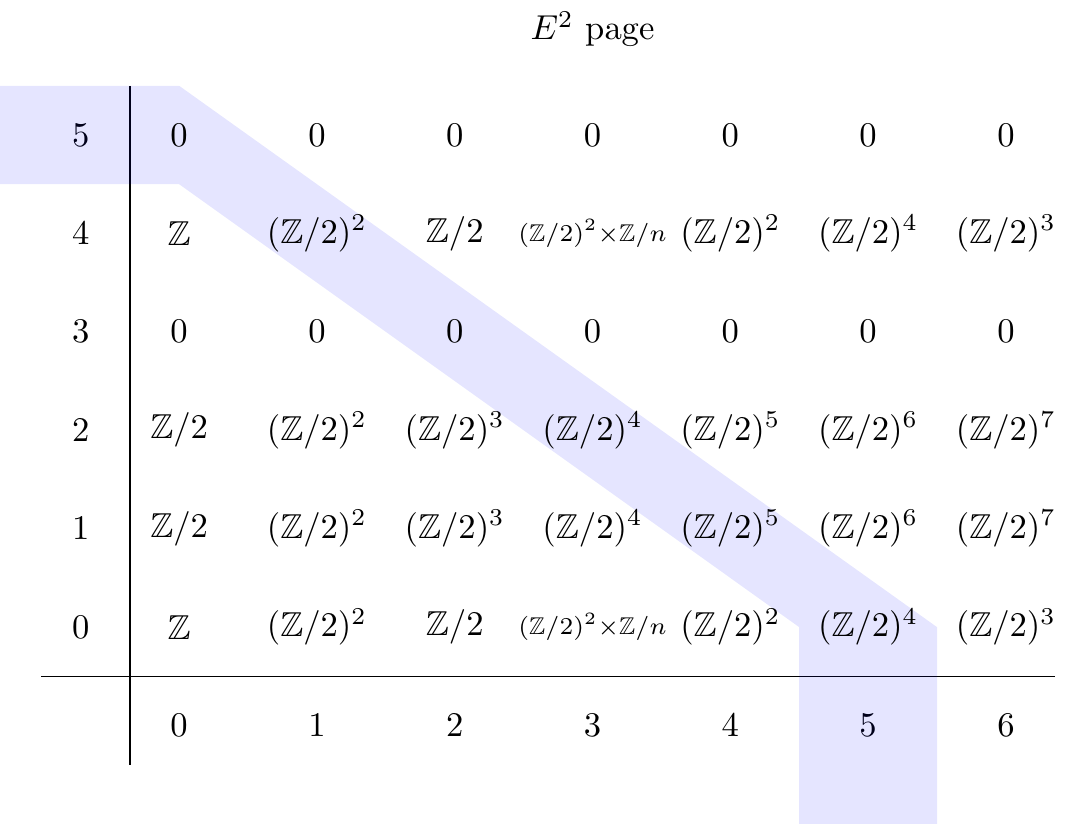}
  \caption{The $E^2$ page of the Atiyah--Hirzebruch spectral sequence for calculating $\Omega^\Spin_d(BD_{2n})$ when $n$ is even.}
  \label{fig:AHSS-BD2evenE2}
\end{figure}

One can do better for $n=2^{k+1}$ and find also the bordism groups in lower degrees. We proceed by first making use of the isomorphism between
$\tilde{\Omega}^{\Spin}_d(BD_{2n})$ and $\widetilde{ko}_d(BD_{2n})$ and note that the K-theory computation has been carried out using the Bockstein spectral sequence in Ref. \cite{bruner2010connective}, which we reproduce in Table \ref{tab:reduced-ko-BD}, where the power $c$ are determined up to a range:  $3 \leq c \leq 2^k+5$.
\begin{table}[h]
  \centering
  \begin{tabular}{|c|ccccc|}
    \hline
    $d$ & $2$ & $3$ & $4$ & $5$ & $6$\\
    \hline
    $\widetilde{ko}_d(BD_{2^{k+2}})$ & $(\Z/2)^c$ & $|\widetilde{ko}_3(BD_{2^{k+2}})|=2^{k+8}$ & $(\Z/2)^2$ & $0$ & $\Z/2$\\
    \hline
  \end{tabular}
  \caption{The reduced $ko$-homology of $BD_{2^{k+2}}$ from degree $2$ through $6$}
  \label{tab:reduced-ko-BD}
\end{table}

The first thing we notice from the table above is that
\begin{equation}
\tilde{\Omega}^{\Spin}_5(BD_{2^{k+2}}) \cong \widetilde{ko}_5(BD_{2^{k+2}}) = 0,
\end{equation}
agreeing with our calculation above. Moreover, the K-theory results also resolves the extension problem in the column $t-s=2$ in the ASS shown in Fig. \ref{fig:AdamsSS-BD2n}, giving us
\begin{equation}
\tilde{\Omega}^{\Spin}_2(BD_{2^{k+2}}) \cong \Z/2 \times \Z/2 \times \Z/2 .
\end{equation}

The only remaining spin bordism groups of $BD_{2^{k+2}}$ in degrees lower than $5$ to be worked out are $\tilde{\Omega}^{\Spin}_3(BD_{2^{k+2}})$ and $\tilde{\Omega}^{\Spin}_4(BD_{2^{k+2}})$. The connective KO-theory computation in Ref. \cite{bruner2010connective}, done via the Bockstein spectral sequence, only gives the order of the group: $\abs{\Omega^{\Spin}_3(BD_{2^{k+2}})}=2^{k+8}$. Fortunately, we can use this information to deduce the differentials from the column $t-s =4$ in the ASS of Fig. \ref{fig:AdamsSS-BD2n} and precisely determine the bordism groups, as follows.

\begin{figure}[h]
  \centering
  \includegraphics[scale=0.6]{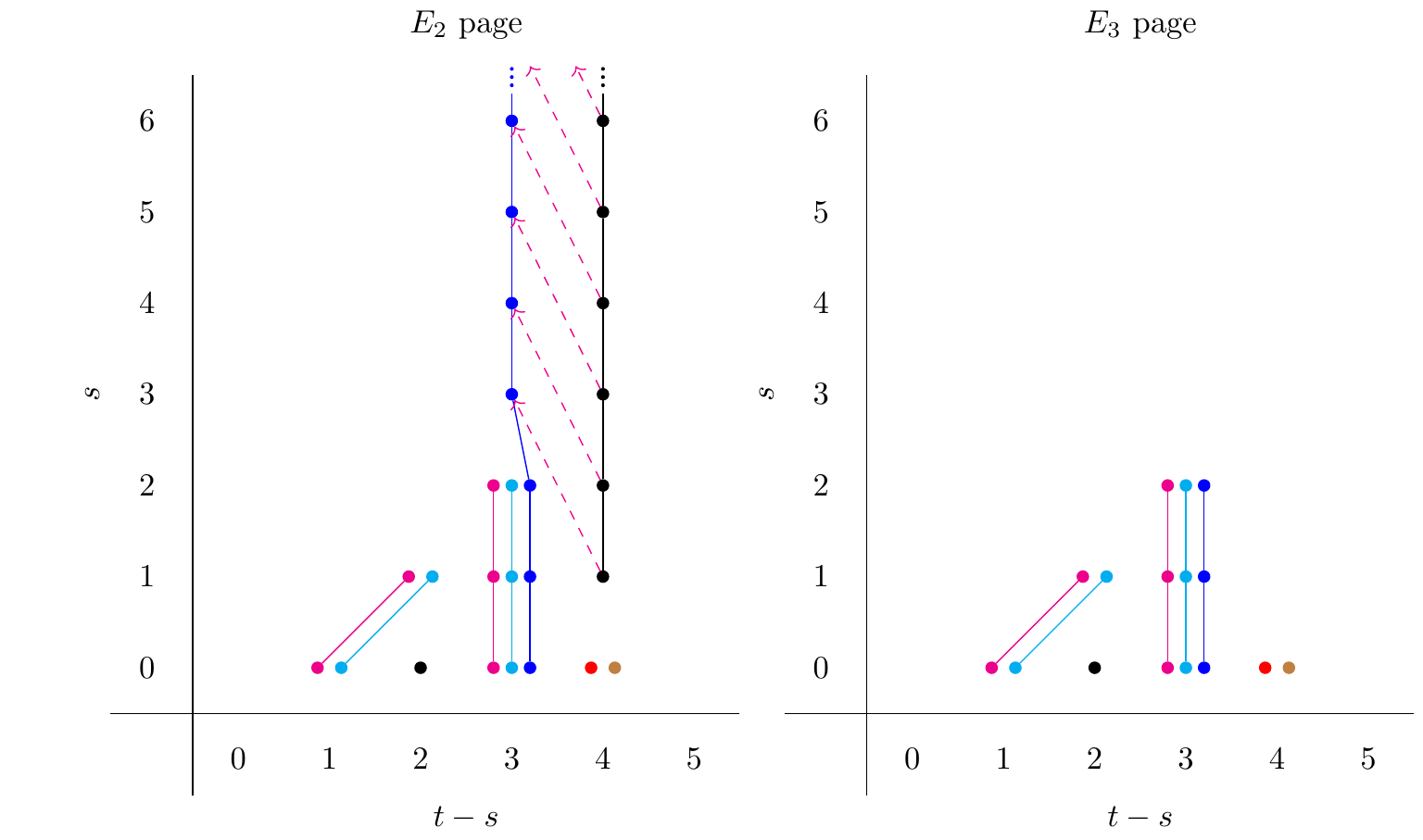}
  \caption{Adams chart for $\ext^{s,t}_{\mathcal{A}(1)}\left(\tilde{H}^\bullet(BD_8;\Z/2),\Z/2\right)$}
  \label{fig:Adams-SS-D8}
\end{figure}

From the computations of the connective real K-theory for $BD_{2\cdot 2^{k+1}}$ in Ref. \cite{bruner2010connective}, we know that
\begin{equation}
  \tilde{\Omega}^{\Spin}_4(BD_{2^{k+2}})\cong \Z/2 \times \Z/2.
\end{equation}
Therefore, there is bound to be a series of differentials $d_r$ on the $r$\textsuperscript{th} page out of the $h_0$-tower in the $t-s=4$ column, where $r$ is still unknown. The differentials will hit part of the $h_0$-tower in the $t-s=3$ column, truncating the extension down to $\Z/2^{r+1}$. The third spin bordism group then has the order $2^{3+3+r+1}=2^{r+7}$. Again, we know from Ref.~\cite{bruner2010connective} that the third spin bordism group of $BD_{2^{k+2}}$ is $2^{k+8}$. Thus, we deduce that
\begin{equation}
r = k+1. \nonumber
\end{equation}
For example, the Adams chart for
$\ext^{s,t}_{\mathcal{A}(1)}\left(\tilde{H}^\bullet(BD_8;\Z/2)\right)$ is shown in
Fig. \ref{fig:Adams-SS-D8}.
We can then conclude that
\begin{equation}
\tilde{\Omega}^{\Spin}_3(BD_{2^{k+2}}) \cong (\Z/8)^2 \times \Z/2^{k+2} .
\end{equation}
These results are summarised in Table \ref{tab:bord-results}.

\section{TQFTs for $Q_8$ structure}
\label{app:tqft-Q8}

In \S \ref{sec:Q8-4d} it was shown that there is no global
anomaly from fermions in any representation of $Q_8$, because the
$\eta$-invariant vanishes identically when evaluated on a generator of $\Omega^{\Spin}_5(BQ_8)$
\cite{Botvinnik:1995a}. 
We might then be tempted to conclude that the
factors $\Z/2 \times \Z/2$ in $\Omega^{\Spin}_5(BQ_8)$ must be associated with
`pure bosonic anomalies' that descend from mod 2 cohomology. Indeed,
the mod 2 cohomology ring of $BQ_8$ is~\cite{Bayen:1996}
\begin{equation}
  \label{eq:mod2-cohomology-BQ8}
H^{\bullet}(BQ_8,\Z/2) \cong \frac{\Z/2 [x,y,w]}{(x^2+xy+y^2, x^2y+xy^2)},
\end{equation}
where $\text{deg}(x)=\text{deg}(y)=1$ while $\text{deg}(w)=4$, and so there are exactly two elements in the degree 5 cohomology group, $xw$ and $yw$. 
However, it does not follow that these cohomology classes are the cobordism generators that we are looking for. It turns out
that the cobordism generators of the $\Z/2\times \Z/2$ depend on the spin structure.

\begin{figure}[h]
  \centering
  \includegraphics[scale=0.7]{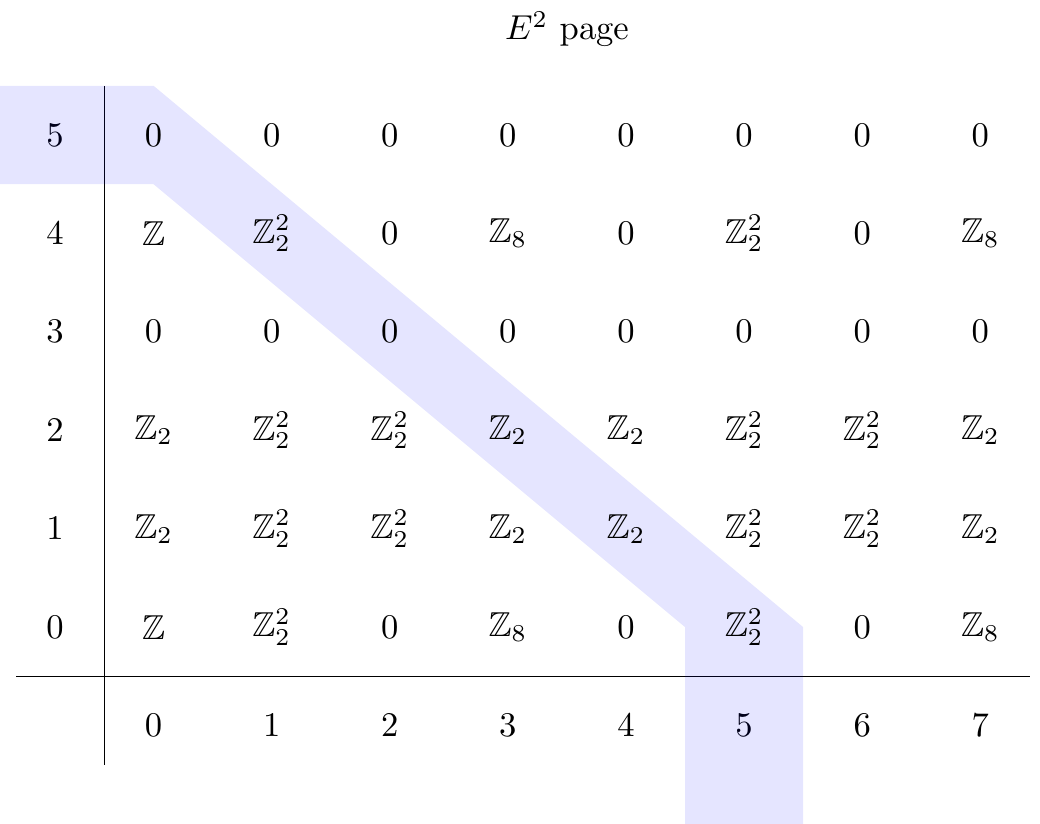}
  \caption{The $E^2$ page of the AHSS for calculating $\Omega^\Spin_\bullet(BQ_8)$}
  \label{fig:AHSS-Q8}
\end{figure}

\begin{figure}[h]
  \centering
  \includegraphics[scale=0.7]{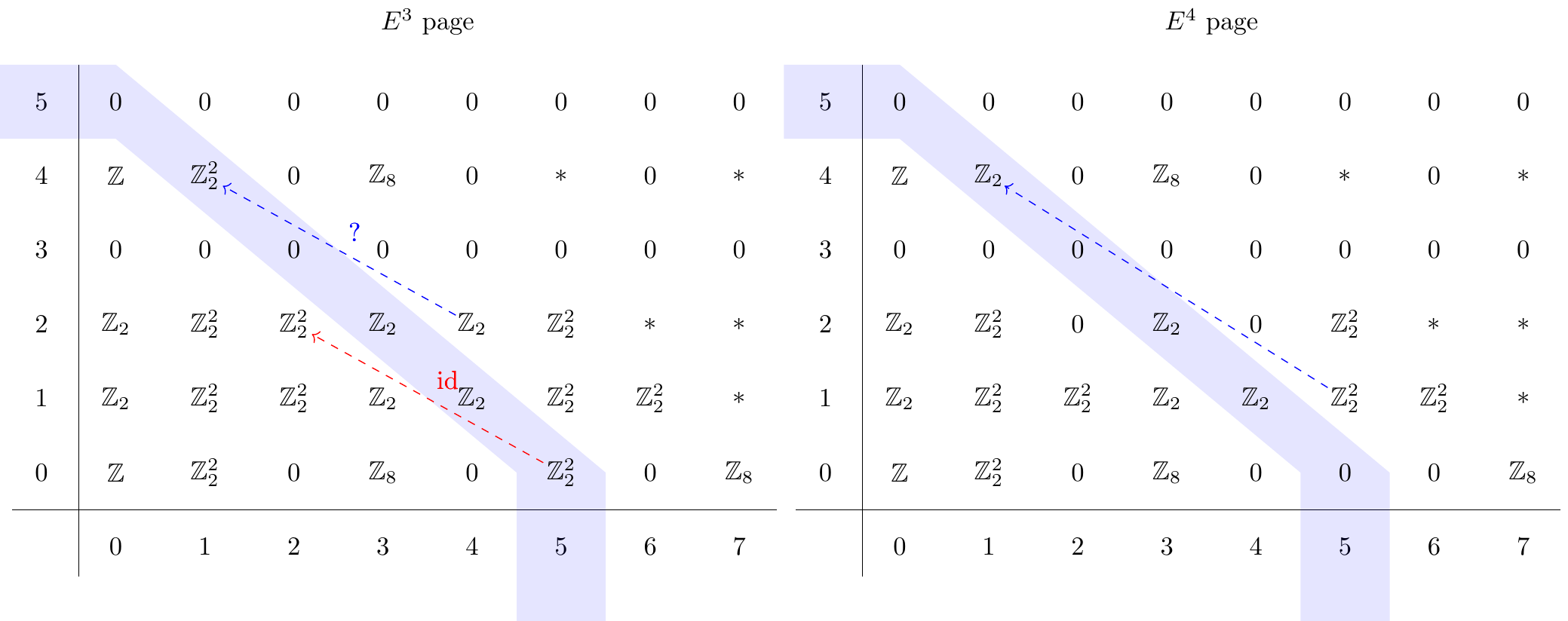}
  \caption{The $E^3$ and $E^4$ pages of the AHSS for $\Omega^\Spin_\bullet(BQ_8)$}
  \label{fig:AHSS-Q8-E3}
\end{figure}

One way to see this is to look at the Atiyah--Hirzebruch spectral
sequence
\begin{equation}
E^2_{p,q} = H_p(BQ_8; \Omega^{\Spin}_q(\text{pt})) \Rightarrow \Omega^{\Spin}_{p+q}(BQ_8)
\end{equation}
for the fibration $\text{pt}\to BQ_8\to BQ_8$, as shown in
Fig. \ref{fig:AHSS-Q8}. Since $\sq^2$ acts trivially on
every generator of $H^{\bullet}(BQ_8;\Z/2)$, it follows that all
differentials on the $E^2$ page that act on the
$0$\textsuperscript{th} and $1$\textsuperscript{st} rows must also be
trivial. To work out some differentials on the $E^3$ page, we compare
the AHSS with known results. Now, we know from the K-theory
computation that $\tilde{\Omega}^{\Spin}_4(BQ_8)\cong \Z/2$. On the
$p+q=4$ diagonal, there are entries $E^3_{2,2}\cong (\Z/2)^2$ and
$E^3_{3,1}\cong \Z/2$. Out of these two, the entry $E^3_{3,1}$ already
stabilises since there are no more non-trivial differentials into or
emanating from it. Thus, it is this entry that gives the resulting
$\Z/2$ factor in $\tilde{\Omega}^{\Spin}_4(BQ_8)$, and the entry
$E^3_{2,2}$ must be killed by the differential
\begin{equation}
d_3: E^3_{5,0}\cong (\Z/2)^2 \to E^3_{2,2} \cong (\Z/2)^2
\end{equation}
as shown in Fig. \ref{fig:AHSS-Q8-E3}.

\begin{figure}[h]
  \centering
  \includegraphics[scale=0.7]{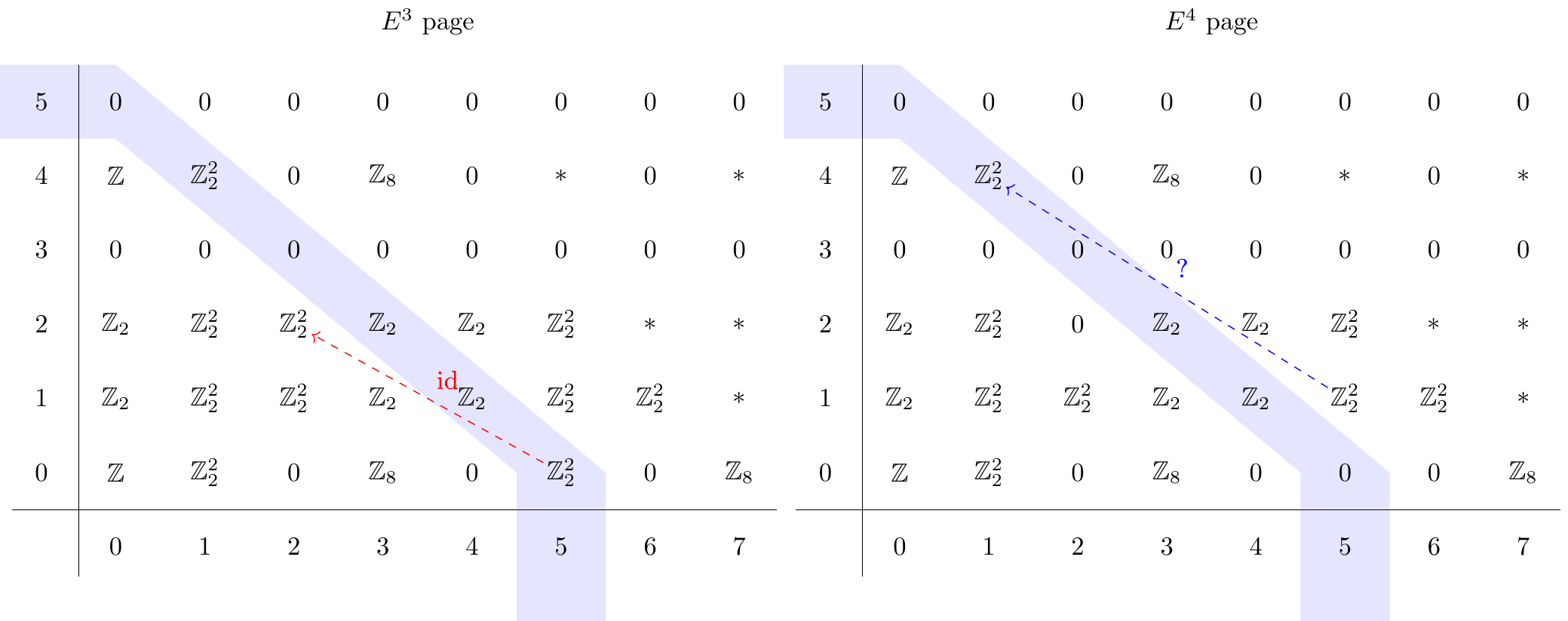}
  \caption{Another possibility for the elimination of the entry $E_{1,4}$}
  \label{fig:AHSS-Q8-extra}
\end{figure}

The $\Z/2$ factors from the entries $E^3_{4,1}$ and $E^3_{3,2}$ also
stabilise at this page. Since $\Omega^{\Spin}_5(BQ_8)\cong (\Z/2)^2$, we can
deduce that the entry $E^3_{1,4}$ must be completely killed by
differentials, either successively from $E^3_{4,2}$ and $E^4_{5,1}$
(which is the situation shown in Fig. \ref{fig:AHSS-Q8-E3}), or in one
go by $d_4$ from $E^4_{5,1}$, as shown in
Fig. \ref{fig:AHSS-Q8-extra}. We can then clearly see that the entries
contributing to the $\Z/2$ factors in $\Omega^{\Spin}_5(BQ_8)$ depends on
the $\Z/2$ factors in $\Omega^{\Spin}_1(\text{pt})$ and
$\Omega^{\Spin}_2(\text{pt})$, which need the existence of the spin
structure as claimed.

The $\Z/2$ in the entry $E^{\infty}_{3,2}$ from AHSS computation above
seems to suggest that the bordism generator for this $\Z/2$ factor is
a product of the torus with periodic spin structure around both
1-cycles and a 3-manifold that is a generator of the $\Z/8$ factor in
$\Omega^{\Spin}_3(BQ_8)$. Indeed, this observation is correct, but we need
to examine the Adams spectral sequence to confirm it.

The Adams spectral sequence (ASS) for our purpose reads
\begin{equation}
E^{s,t}_2 = \ext^{s,t}_{\mathcal{A}(1)}\left( H^{\bullet}(BQ_8;\Z/2),\Z/2 \right) \Rightarrow \Omega^{\Spin}_{t-s}(BQ_8)^{\wedge}_2,
\end{equation}
with the input being the mod 2 cohomology ring of $BQ_8$, given by Eq. \eqref{eq:mod2-cohomology-BQ8}.
Using \texttt{GAP} \cite{GAP4}, the action of $\mathcal{A}(1)$ on the generators of $H^{\bullet}(BQ_8;\Z/2)$ that cannot be obtained from the Axioms reads
\begin{equation}
\sq^1 w =0, \qquad \sq^2 w =0.
\end{equation}
This information is enough to determine the structure of
$H^{\bullet}(BQ_8;\Z/2)$ as an $\mathcal{A}(1)$-module, which we present in
Fig. \ref{fig:Q8-ASS-cells} below.
\begin{figure}[h]
  \centering
  \includegraphics[scale=0.6]{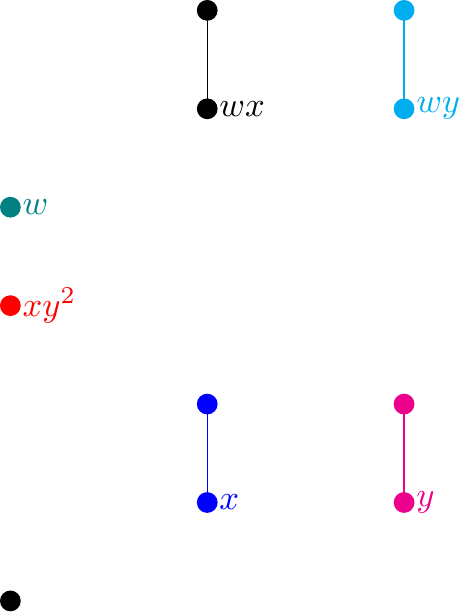}
  \caption{The $\mathcal{A}(1)$-module structure of $H^{\bullet}(BQ_8;\Z/2)$ up to degree $5$}
  \label{fig:Q8-ASS-cells}
\end{figure}
Algebraically, we can write
\begin{equation}
H^{\bullet}(BQ_8;\Z/2) \cong \Z/2 \oplus \Sigma S \oplus \Sigma S \oplus \Sigma^3\Z/2 \oplus \Sigma^4\Z/2 \oplus \Sigma^5 S \oplus \Sigma ^5S\oplus \ldots ,
\end{equation}
where $S$ is the ``stick'' $\mathcal{A}(1)$-module. Since we already
know the Adams chart for $\ext^{s,t}_{\mathcal{A}(1)}(\Z/2,\Z/2)$, we
only need calculate the Adams chart for
$\ext^{s,t}_{\mathcal{A}(1)}(S,\Z/2)$ to obtain the $E_2$ page of the
ASS for $H^{\bullet}(BQ_8;\Z/2)$.

\begin{figure}[h]
  \centering
  \includegraphics[scale=0.7]{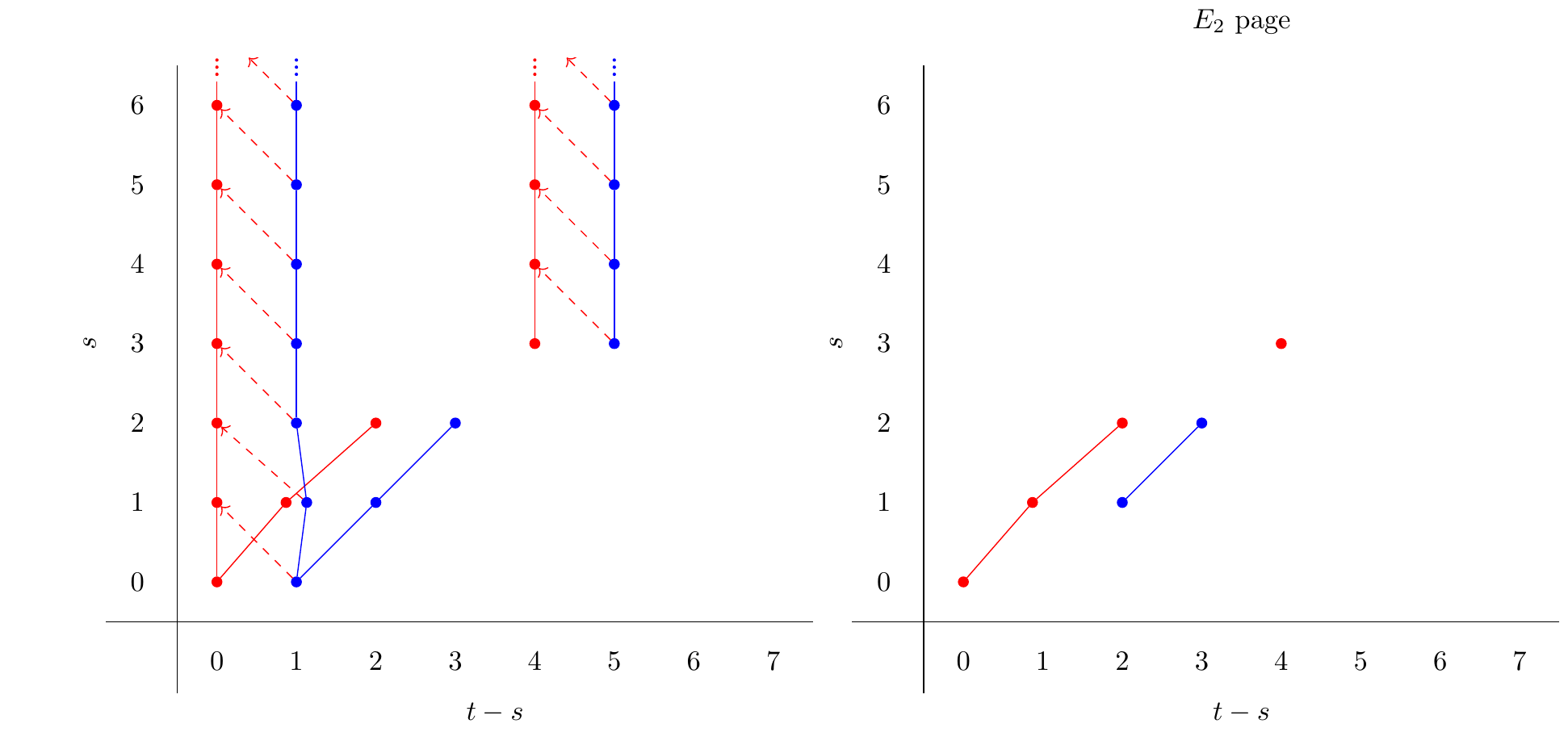}
  \caption{The Adams chart for $\ext^{s,t}_{\mathcal{A}(1)}(S,\Z/2)$}
  \label{fig:Adams-chart-stick}
\end{figure}

Again, we put the module $S$ in the middle of a short exact sequence, which reads
\begin{equation}
0 \to \Sigma \Z/2 \to S \to \Z/2 \to 0,
\end{equation}
which pictorially looks like
\begin{equation}
\includegraphics[scale=0.6]{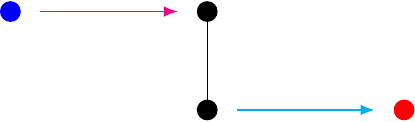}\nn
\end{equation}
This short exact sequence induces a long exact sequence in the
$\ext$-functor \cite{beaudry2018guide}, whose connecting morphism
\begin{equation}
\delta: \ext^{s,t}_{\mathcal{A}(1)}(\Sigma\Z/2,\Z/2) \to \ext^{s,t}_{\mathcal{A}(1)}(\Z/2,\Z/2),
\end{equation}
which translates into a differentials of slope $(-1,1)$ in the Adams
chart where the charts for $\Sigma\Z/2$ and $\Z/2$ are drawn together as
shown in Fig. \ref{fig:Adams-chart-stick}. We thus obtain the Adams
chart for the stick module $S$ as shown on the right of the figure.

\begin{figure}[h]
  \centering
  \includegraphics[scale=0.55]{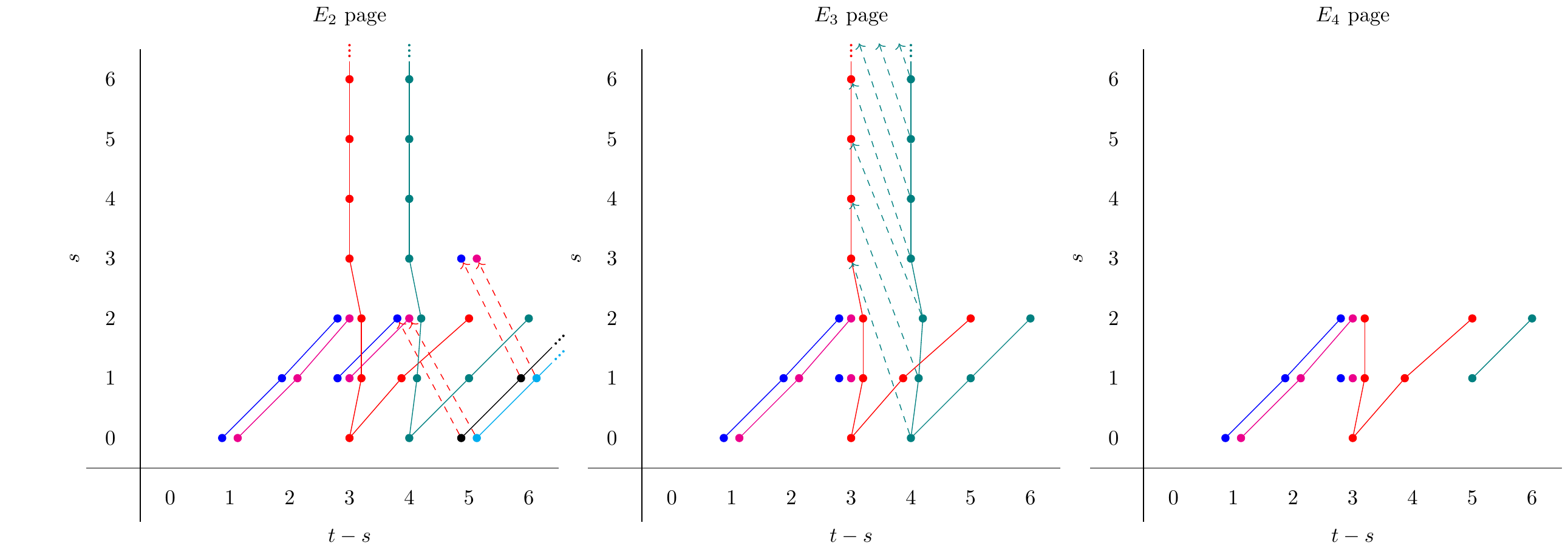}
  \caption{The Adams spectral sequence calculating the reduced spin bordism groups of $BQ_8$}
  \label{fig:ASS-BQ8}
\end{figure}

We now combine the Adams chart of $S$ with the Adams chart of $\Z/2$, and
ignoring the $\Z/2$ cell at the bottom that only gives us the factors
corresponding to $\Omega^{\Spin}_{\bullet}(\text{pt})$. Then, the Adams charts for the
$E_2$, $E_3$, $E_4$ pages of the Adams spectral sequence that computes
the reduced spin bordism groups of $BQ_8$ is shown in
Fig.~\ref{fig:ASS-BQ8}. The differentials are deduced in such a way
that the resulting bordism groups are compatible with the results from
K-theory computation in \cite{Bayen:1996}.

In the Adams chart, the vertical line indicates the non-trivial
extension by $\Z/2$ and the slanted line of slope 1 indicates a
Cartesian product by a circle with periodic spin structure.\footnote{We thank Arun Debray for helpful discussions about this.} In the
final page of Fig. \ref{fig:ASS-BQ8}, we see that one $\Z/2$ factor in
$\Omega^{\Spin}_5(BQ_8)$ is at the end of a sequence of two slanted lines
starting at the generator of the $\Z/8$ factor in
$\Omega^{\Spin}_3(BQ_8)$. \footnote{We know that the element at the
  beginning of the sequence of two slanted lines generates $\Z/8$
  because it is also a starting point of two successive vertical
  lines.} Ref. \cite{Botvinnik:1995a} showed that a generator of the $\Z/8$ factor can be taken to be the quotient $S^3/\tau(Q_8)$, where $\tau$ is the 2-dimensional representation of $Q_8$, which
can be thought of as a map $\tau:Q_8 \to \SU(2)\cong S^3$ given explicitly by
\begin{equation}
\tau(\pm 1) = \pm \mathbb{I}_2, \quad \tau(\pm i) = \pm \ii \sigma_3, \quad \tau(\pm j) = \pm \ii \sigma_2, \quad \tau(\pm k) = \pm \ii \sigma_1,
\end{equation}
$\sigma_i$ being the usual Pauli matrices. Thus, one of the two bordism generators of
$\Omega^{\Spin}_5(BQ_8)$ is
\begin{equation}
T^2_{PP} \times S^3\big/ \tau(Q_8),
\end{equation}
as hinted in the AHSS, where $T^2_{PP}$ is a torus with periodic spin
structure around both 1-cycles. The corresponding bordism invariant is
\begin{equation}
  \text{Arf}\cup xy^2,
\end{equation}
where $\text{Arf}$ is the Arf invariant of the spin structure.

Notice that the cohomology classes $xw$ and $yw$, identified at the start of this Appendix, are trivial on this generator of the bordism group: we can factorise {\em e.g.} $xw(T^2_{PP} \times S^3\big/\tau(Q_8)) = x(S^1_P) w(S^1_P \times S^3\big/\tau(Q_8))$ which vanishes because the $S^1$ does not come with a non-trivial $Q_8$ bundle, and so the cohomology class $x \in H^1(BQ_8; \Z/2)$ is trivial when evaluated on it.\footnote{To see this, note that  $x(S^1_P)$ is shorthand for the pairing $\left\langle f^{*}(x),S^1_p \right\rangle$, where $f^{*}:H^{\bullet}(BQ_8;\Z/2) \to H^{\bullet}(S^1_P;\Z/2)$ is the pullback of the classifying map $f:S^1_P\to BQ_8$. Since the $Q_8$-bundle on $S^1_P$ is trivial, $f$ is trivial, and so does $f^{*}(x)$.} The same argument works for $yw$. While not giving an explicit calculation, we expect a similar story to play out for the second independent bordism invariant. 

\section{Bordism generators of $\Omega^{\Spin}_5(B\SL(2,\mathbb{F}_3))$ and $\Omega^{\Spin}_5(BQ_8)$}
\label{app:bord-gen-SL23}

In this Appendix, we will argue that the manifolds generating the factor $\Z/2\times \Z/2$ of $\Omega^{\Spin}_5(B\SL(2,\mathbb{F}_3))$ are also generators of $\Omega^{\Spin}_5(BQ_8)\cong \Z/2 \times \Z/2$.

Since we are only concerned with a mod 2 effect, we can work at $2$-completion without loss of generality. A generator of the $\Z/2 \times \Z/2$ is given by a spin 5-manifold $X$ together with a map
\begin{equation}
f : X \to B\SL(2,\mathbb{F}_3).
\end{equation}
Note that the subgroup embedding $Q_8 \hookrightarrow \SL(2,\mathbb{F}_3)$ induces a map $i:BQ_8 \to B\SL(2,\mathbb{F}_3)$. Moreover, at 2-completion, there is a stable decomposition \cite{MARTINO199113}
\begin{equation}
BQ_8 \cong B\SL(2,\mathbb{F}_3) \vee \Sigma^{-1}\left(BS^3/BN \right)\vee \Sigma^{-1}\left(BS^3/BN\right),
\end{equation}
where $N$ is the normaliser of a maximal torus in $S^3$, and $\vee$ is the wedge sum. Thus, the map $f$ naturally induces a map $f^{\prime}:X\to BQ_8$ such that $i\circ f^{\prime} = f$ and we can also think of $X$ as equipped with a $Q_8$-bundle.  Now, suppose that $(X,f^{\prime})$ is nullbordant, that is, there exists a spin 6-manifold $Y$ with a map $g:Y\to BQ_8$ such that $\partial Y = X$ and $g|_\partial = f$. Using the embedding $Q\hookrightarrow \SL(2,\mathbb{F}_3)$, we now have a spin 6-manifold $Y$ whose boundary is $X$ with a map $i\circ g:Y \to B\SL(2,\mathbb{F}_3)$ such that $i\circ g|_{\partial} = i \circ f^{\prime} = f$. Hence, $(X,f)$ is nullbordant, which is a contradiction. Therefore, manifolds that generate the $\Z/2 \times \Z/2$ factor in $\Omega^{\Spin}_5(B\SL(2,\mathbb{F}_3))$ also generate $\Omega^{\Spin}_5(BQ_8)$.

\section{Decomposition of Standard Model representation}
\label{app:SM}

In this short Appendix we derive the decomposition \eqref{eq:SM-decomp} used in our proof of the anomaly structure of the group $\SL(2,\F_3)$. It can be deduced via character theory as follows. First, note that any element of the group $\UU(2)$ is conjugate to a diagonal matrix of the form $\diag \left( \ee^{\ii \theta_1},\ee^{\ii \theta_{2}} \right)$, so we can label a conjugacy class by a pair of angles $(\theta_1,\theta_2)$. The character of the conjugacy class $(\theta_1,\theta_2)$ in the irrep $(\textbf{n}, q)$ is given by the \emph{Weyl character formula}
\begin{equation}
  \label{eq:Weyl-char-formula-U2}
  \chi_{n,k}(\theta_1,\theta_2) = \ee^{\ii (q-n+1) \frac{\theta_1+\theta_2}{2}} \left[ \frac{\ee^{\ii n \theta_1}-\ee^{\ii n \theta_2}}{\ee^{\ii \theta_1}-\ee^{\ii \theta_2}} \right] .
\end{equation}
Each conjugacy class of the subgroup $\SL(2,\mathbb{F}_3)$ under the subgroup embedding \eqref{eq:SL23-U2-embedding} can now be alternatively labelled by $(\theta_1,\theta_2)$, given in Table \ref{tab:SL2F-U2-char}, with the conjugacy classes arranged in the same order as the character table of $\SL(2,\mathbb{F}_3)$ (Table \ref{tab:SL2F-char}).  Applying the Weyl character formula above to each of these conjugacy classes, we arrive at the character table of the $\UU(2)$ irrep $(\textbf{n},q)$ restricted to the subgroup $\SL(2,\mathbb{F}_3)$. The result is again tabulated in Table \ref{tab:SL2F-U2-char}. We can now use the orthogonality of the characters to obtain the decomposition for each irrep $(\textbf{n},q)$. Applying the outlined procedure to the irreps in the Standard Model representation \eqref{eq:SM-rep}, we easily obtain the decomposition \eqref{eq:SM-decomp} as claimed.
\begin{table}[h]
  \centering
  \begin{tabular}{c|ccccccc}
    & $1$ & $1$ & $4$ & $4$ & $6$ & $4$ & $4$\\
    $(\theta_1,\theta_2)$ & $(0,0)$ & $(\pi,-\pi)$ & $(0,-\frac{2\pi}{3})$ & $(0,\frac{2\pi}{3})$ & $(\frac{\pi}{2},-\frac{\pi}{2})$ & $(\pi,\frac{\pi}{3})$ & $(-\pi,-\frac{\pi}{3})$\\
    \hline
    &&&&&&&\\
    $(\textbf{n},q)$ & $n$ & $n(-1)^{n-1}$ & $\ee^{-\frac{\pi\ii q}{3}}\frac{\sin \frac{n\pi}{3}}{\sin \frac{\pi}{3}}$ & $\ee^{\frac{\pi\ii q}{3}}\frac{\sin \frac{n\pi}{3}}{\sin \frac{\pi}{3}}$  & $\sin \frac{n\pi}{2}$ & $\ee^{\frac{2\pi\ii q}{3}}\frac{\sin \frac{n\pi}{3}}{\sin \frac{\pi}{3}}$  & $\ee^{\frac{-2\pi\ii q}{3}}\frac{\sin \frac{n\pi}{3}}{\sin \frac{\pi}{3}}$ \\
  \end{tabular}
  \caption{Characters for the irreps of $\UU(2)$ restricted to the subgroup $\SL(2,\mathbb{F}_3)$ with the subgroup embedding \eqref{eq:SL23-U2-embedding}.}
  \label{tab:SL2F-U2-char}
\end{table}

\bibliographystyle{JHEP} 
\bibliography{references}
\end{document}